\documentclass[letterpaper,twocolumn,10pt]{article}
\usepackage{usenix}

\usepackage{tikz}
\usepackage{amsmath}
\usepackage{booktabs}
\usepackage{comment}
\usepackage[table]{xcolor}
\definecolor{dark green}{RGB}{94,145,40}

\newcommand\mypara[1]{\noindent \textbf{#1}}
\usepackage{xcolor}

\newif\ifshownotes
\shownotestrue 
\definecolor{darkgreen}{RGB}{0,100,0}
\newcommand{\note}[3]{%
  \ifshownotes
    \textcolor{#1}{\emph{#2 --#3}}%
  \fi
}

\newcommand{\mlnote}[1]{\note{orange}{#1}{ML}}

\providecommand{\pd}{Pump-and-Dump}
\providecommand{\bt}{Bait-and-Trap}
\providecommand{\ie}{\emph{i.e.,} }
\providecommand{\eg}{\emph{e.g.,} }
\providecommand{\tool}{\textsc{HoneyCheck}}

\begin{document}

\date{}

\title{\pd{} meets Honeypot Tokens: \\  Detection and Analysis of Telegram \bt{} Schemes}


\author{
{\rm Federico Cernera}\\
\text{cernera@di.uniroma1.it}\\
Sapienza University of Rome 
\and
{\rm Massimo La Morgia}\\
\text{lamorgia@di.uniroma1.it}\\
Sapienza University of Rome
\and 
{\rm Alessandro Mei}\\
\text{mei@di.uniroma1.it}\\
Sapienza University of Rome
\and
{\rm Francesco Sassi}\\
\text{sassi@di.uniroma1.it}\\
Sapienza University of Rome
}
\maketitle

\begin{abstract}
While \pd{} schemes have been extensively studied on centralized exchanges (CEXs), how they operate in decentralized exchanges (DEXs) remains largely unexplored. 
We monitor 83 Telegram channels used to coordinate \pd{} campaigns and collect 3,677 events across the BNB Smart Chain and Ethereum. 
Our analysis reveals that, although these operations superficially resemble CEX-based \pd{} schemes, their underlying mechanism is fundamentally different. Rather than manipulating prices, organizers orchestrate deceptive \pd{} campaigns around honeypot tokens whose smart contracts allow Telegram subscribers to buy but prevent them from selling. Unaware of this restriction, victims purchase these tokens and are unable to recover their funds. We refer to this new fraud mechanism as \emph{\bt{}}.
We show that \bt{} provides a substantially more reliable profit strategy than \pd{} schemes: organizers gain in 99.3\% of cases, extracting over \$7 million in profit. 
To counter this threat, we develop a transaction-simulation tool that detects honeypot tokens before purchase by testing whether a user can successfully sell against the token's live contract state, providing a practical defense against \bt{} operations.

\end{abstract}

\section{Introduction}

\pd{} schemes are market manipulations in which a group of people buys a low-priced asset, artificially inflates its price, and then sells their holdings to unsuspecting investors~\cite{allen1992stock, huang2015stock}.
Although the phenomenon first emerged in stock markets, it has recently become increasingly prevalent in cryptocurrency exchanges~\cite{la2023doge,xu2019anatomy,kamps2018moon,hamrick2019economics}.
The scheme is coordinated through public Telegram~\cite{telegram_desktop} channels.
At a predetermined time, the organizer announces a target cryptocurrency, which participants buy as quickly as possible to drive up its price.
The goal is to attract outside investors, who mistake the price increase for genuine market demand.
Then, participants sell their previously acquired cryptocurrencies and, if they exit before the subsequent price collapse, realize a profit.
While prior work has focused almost exclusively on centralized exchanges (CEXs)~\cite{la2023doge,xu2019anatomy,kamps2018moon,hamrick2019economics,mahrous2026pumpsense}, the rapid growth of decentralized exchanges (DEXs) has introduced a new trading paradigm.
It remains largely unexplored whether \pd{} schemes also occur in this ecosystem, and whether they exhibit the same characteristics as on CEXs~\cite{li2021cryptocurrency,rath2024profiling}.


In this paper, we analyze 83 Telegram channels that 
perform
3,677 operations advertised as \pd{} schemes on DEXs across the BNB Smart Chain (BSC) and Ethereum. Channel administrators announce a target token and encourage subscribers to purchase it quickly, with the expectation of profiting from a rapid price increase. The corresponding on-chain execution, however, reveals a fundamentally different mechanism. In most of the observed operations, the target is a \emph{honeypot token}: its smart contract allows ordinary addresses to purchase the token but prevents them from selling it, while a small set of addresses, which we refer to as \emph{colluding addresses}, can trade freely. The advertised \pd{} thus provides the bait, while the honeypot restrictions embedded in the token contract provide the trap. We term this phenomenon \emph{\bt{}}.

To identify honeypot tokens, we develop and publicly release \tool{}. For each operation, \tool{} emulates a token purchase followed by an attempted sale on a fork reproducing the blockchain state at the time of the operation. Using this approach, we classify 2,629 operations as \bt{}, accounting for 71.5\% of our dataset. \bt{} is therefore the dominant phenomenon in our dataset, suggesting that it constitutes a common form of fraud among Telegram-coordinated operations advertised as \pd{} schemes on DEXs.
Then, we identify colluding addresses and find that, although they represent only 10.6\% of the addresses participating in \bt{} operations, they account for more than 72\% of the total buy volume and 92\% of the total sell volume. By generating most of the observed trading activity, these addresses create the appearance of a conventional \pd{} operation while concealing the trading restrictions imposed on regular participants.
We then quantify the economic impact of \bt{} and find that token creators and their associated colluding addresses realize positive returns in 99.3\% of the analyzed cases, earning more than \$7 million in aggregate. Given the resulting economic harm, detecting honeypot behavior before a purchase is made is critical to preventing further losses. By applying the same transaction-simulation procedure used for historical analysis to the current blockchain state, \tool{} enables investors to determine within seconds whether a buy-and-sell sequence would complete successfully before committing any funds.


\begin{itemize}

\item \textbf{Uncovering \bt{} schemes}. By jointly analyzing Telegram coordination and on-chain execution across 3,677 events, we uncover a distinct fraud behind operations advertised as \pd{}. 
We term this fraud \bt{} and find that it accounts for 71.5\% of the observed operations, making it the dominant phenomenon in our dataset.

\item \textbf{Characterizing \bt{} schemes.} We characterize the lifecycle, coordination, and economic impact of \bt{} operations. We find that 91.1\% of the targeted tokens are deployed in the 48 hours preceding the corresponding operation. We also identify colluding addresses that engage in artificial trading activity, making \bt{} operations resemble conventional \pd{} schemes. Fund-flow tracing further reveals links among apparently unrelated operations, uncovering recurring campaigns financed with the proceeds of earlier operations. Finally, we find that organizers profit from 99.3\% of the operations, extracting more than \$7 million in total.


\item \textbf{\tool{} and Datasets.} We release \tool{}, an open-source tool that emulates token purchases and sales at historical or current blockchain states. We also publicly release, to the best of our knowledge, the largest dataset of announced \pd{} operations on DEXs, comprising 3,677 events and the source code of 1,377 associated token contracts, together with real-world code snippets implementing the honeypot mechanisms we uncover. We believe these resources may support future research on honeypot detection, coordinated fraud, and market manipulation on DEXs.
\end{itemize}


\section{Background}

\subsection{Ethereum and the BNB Smart Chain}
Ethereum is one of the most prominent blockchains.
A key feature of Ethereum is its support for smart contracts, programs executed by the Ethereum Virtual Machine (EVM). Developers write smart contracts in high-level languages such as Solidity~\cite{dannen2017introducing}. The compilation process produces EVM bytecode, which is deployed by sending a transaction to the network. The state of the contract (\eg variable values) is stored in the world state of the blockchain, which is updated by the EVM each time new transactions are added to the blockchain~\cite{wood2018ethereum}.
Smart contracts have enabled the creation of tokens: digital assets issued and managed entirely on-chain, whose supply, transfer rules, and trading logic are defined by the contract code at deployment. The Ethereum Foundation introduced the ERC-20 standard~\cite{erc-20}, a common interface for tokens on the Ethereum blockchain. The standard mandates that compliant smart contracts implement a defined set of functions and events, including \textit{transfer}, \textit{transferFrom} and \textit{approve}. The \textit{transfer} function transfers tokens directly from the caller's address to a recipient. The \textit{transferFrom} function, by contrast, enables a pre-authorized third party, such as a smart contract, to move tokens on behalf of a token holder, provided that the holder has previously granted allowance via the \textit{approve} function. 
This delegated transfer mechanism is essential to decentralized trading, as it allows smart contracts to move tokens on a user's behalf without taking custody of their assets. Critically, the ERC-20 standard mandates the interface but places no restrictions on the underlying implementation. This leaves token creators free to implement arbitrary logic that alters the expected behavior of a token.

\mypara{BNB Smart Chain.}  
The EVM has been adopted by numerous blockchains. One of the most popular is the BNB Smart Chain (BSC)~\cite{paperinu}, launched by Binance in 2020. As an EVM-compatible chain, BSC inherits Ethereum's smart contract architecture and execution model. Token contracts on BSC follow the BEP-20 standard, which mirrors ERC-20 in terms of function signatures and event structures, making contracts largely portable across the two chains.

\subsection{Uniswap and its forks}
\label{sec:background_amm}
Uniswap~\cite{adams2020uniswap} is one of the most widely adopted Automated Market Makers (AMMs). These platforms enable cryptocurrency trading directly on the blockchain without intermediaries. Rather than matching buyers and sellers through an order book, AMMs rely on liquidity pools, smart contracts holding reserves of token pairs, that are traded through swaps. In Uniswap V2 and its forks, these pools follow a constant product formula, whereby the product of the token reserves remains invariant across trades.
Consider a user who wants to swap token $A$ for token $B$ using a liquidity pool consisting of the pair $(A, B)$, where $x$ and $y$ denote the reserve quantities of tokens $A$ and $B$, respectively. The pool enforces the constant product invariant $x \cdot y = k$. When the user deposits an amount $\delta x$ of token $A$, the pool computes the output amount $\delta y$ of token $B$ such that the value of the $k$ invariant  is preserved:$(x + \delta x)(y - \delta y) = k$.

Uniswap V2 implements swaps using two smart contracts: the Pair contract and the Router contract. Pair contracts hold the token reserves and implement the AMM trading logic. In contrast, router contracts serve as an entry point, orchestrating the sequence of contract calls required to execute a swap. Swap operations are classified as buy or sell depending on which Router functions are invoked.

\mypara{Buying swaps.} To buy a token, a user invokes \textit{swapExactETHForTokens} on the Router, attaching the desired amount of ETH to the transaction. The Router transfers the ETH to the Pair contract, which then computes the output by applying the constant product formula. The computed amount of tokens is credited to the user by calling the \textit{transfer} function.

\mypara{Selling swaps.} Selling a token requires an additional preliminary step, as the Router cannot unilaterally withdraw tokens from a user's wallet. The user must first invoke \textit{approve} on the token contract, granting the Router a spending allowance. The user then calls \textit{swapExactTokensForETH} on the Router, which pulls the tokens from the user's wallet into the Pair contract via \textit{transferFrom} and subsequently invokes the Pair's \textit{swap} function. The Pair sends ETH to the Router conditional on having received sufficient input tokens to satisfy the AMM invariant; the Router then converts it into native ETH and forwards it to the user. 

Since Uniswap V2 open-sourced its smart contracts, several protocols have forked its codebase across multiple blockchains~\cite{paperinu}. One of the most prominent is PancakeSwap~\cite{pancakeswap} on BNB Smart Chain. As a direct fork, it implements the same mechanisms described above.

 \begin{figure}[h]   
    \centering
    \includegraphics[scale=0.23]{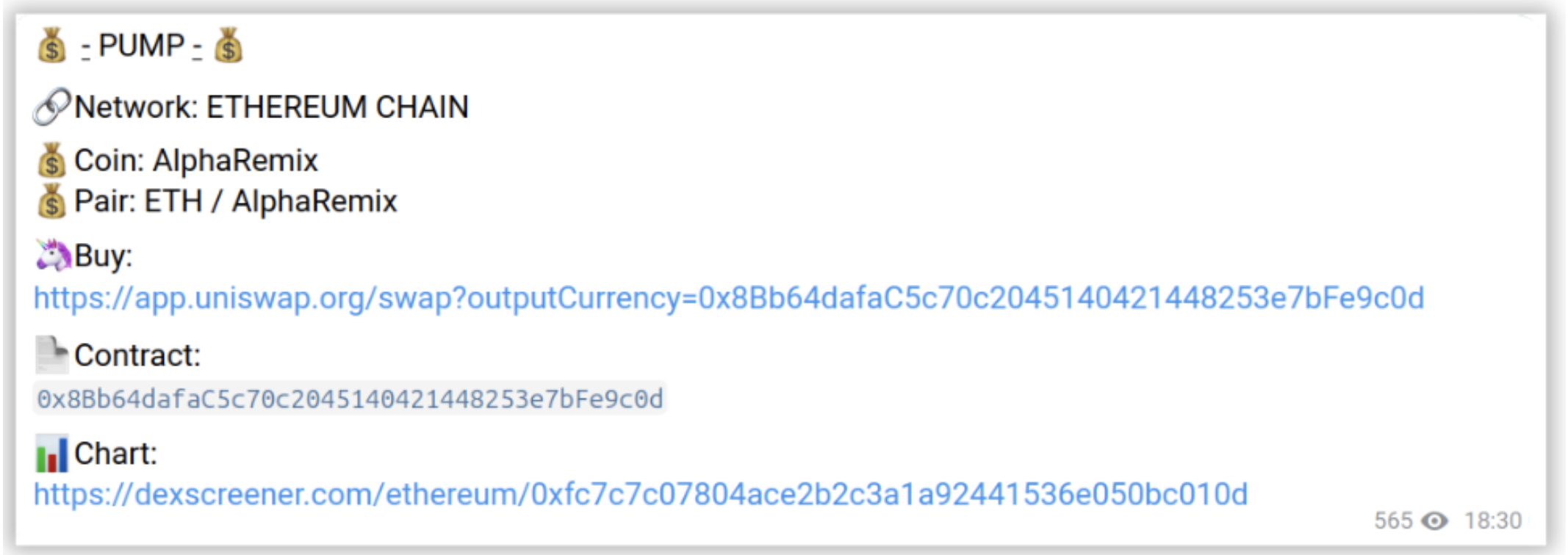}
    \caption{Telegram message signaling the start of a Pump-and-Dump operation.}
    \label{fig:pump_message}
\end{figure}

\section{Dataset}
\label{sec:dataset}

Studying \pd{} schemes organized on Telegram requires combining off-chain coordination data with on-chain trading activity. This section describes how we build our dataset, including the Telegram channels that organize the operations, the announced events, and the complete on-chain transaction history of each target liquidity pool.

\mypara{Channel collection.}
We begin by collecting Telegram channels that organize \pd{} operations on DEXs. 
We rely on a keyword-based search across both the official Telegram Desktop client~\cite{telegram_desktop} and TGStat.com~\cite{tgstat}, a widely used, publicly accessible index of Telegram channels adopted in prior work~\cite{la2023doge, imperati2025conspiracy}. 
Following previous studies~\cite{la2023doge, dhawan2023new}, we compiled a list of broad, generic search terms indicative of \pd{} activity (\eg "pump", "pump and dump", "crypto pump"), reported in Table~\ref{tab:pump_and_dump_keywords} in the Appendix, deliberately avoiding assumptions about specific platforms to maximize coverage.
We conducted this search between November 2024 and January 2025, discovering 183 channels.

\mypara{Channel validation.}
We manually inspected the messages and images shared within each channel, searching for explicit evidence of operations advertised as \pd{} operations targeting DEXs. 
Figure~\ref{fig:pump_message} shows a representative message. 
The message explicitly specifies the target token (\ie AlphaRemix), provides a link to purchase the token on the designated platform (\ie Uniswap), and includes additional details such as the contract address.
While inspecting the channel messages, we further expanded our seed set through snowball sampling. 
Indeed, since keyword search alone cannot surface channels that deliberately avoid such terms, we followed the forwards and mentions linking channels to one another, reaching channels that never appear in keyword results. 
Across keyword search and snowball expansion, we obtained 217 candidate channels, of which we retained the 83 that organized at least one operation targeting a DEX.

\mypara{Pump-and-Dump event extraction.}
\label{subsub:pd_event_extraction}
For each channel, we stored relevant metadata, like the name, description, and creation time.
Two researchers manually inspected the messages posted by each channel from its creation through January 2025 to identify \pd{} operations. This process was facilitated by the high structural regularity of \pd{} announcements, which follow near-identical templates across operations, similar to Figure~\ref{fig:pump_message}, with only token-specific details such as the target token or liquidity pool address varying between them. 
Consequently, annotator disagreements were negligible, and any minor discrepancies were resolved through discussion and consensus.
For each operation, we recorded the start timestamp, the target token, the target liquidity pool, the DEX used, and the blockchain. 
This process resulted in a total of 3,684 operations.
Although our channel discovery process targeted no particular chain or platform, the operations concentrate on the BNB Smart Chain (BSC), with 3,649 operations, and 35 on Ethereum.
We discuss this concentration further in Sec.~\ref{sec:discussion}.
The operations span May 2021 through January 2025.

\mypara{Recovery of missing information.}
\label{subsub:recovery_missing}
In 52.8\% of operations, we extract all relevant information directly from the Telegram messages. 
The remaining cases are incomplete because many channels delete operation details (\ie the target token address, the liquidity pool address, or both) shortly after the operation concludes (Figure~\ref{fig:telegram_message_obfuscated} in the Appendix).
Administrators justify this as preventing latecomers from buying the token after the operation ends. 
However, it also prevents subscribers from independently auditing a channel's past operations. 
The information that is commonly missing is either the address of the target token, the address of the liquidity pool or both.
When only the pool address is known, we query the BscScan~\cite{bscscan_api} and Etherscan~\cite{etherscan_api} APIs to identify the two tokens that form the pool. 
If only the target token address is known, we obtain the pool address by calling the \texttt{getPair()}~\cite{get_pair_function} function of the respective DEX smart contracts. 
Finally, when only the token symbol is provided, we use Moralis~\cite{get_token_by_symbol}, a Web3 development platform, to retrieve all token addresses on BSC or Ethereum that match the symbol~\cite{get_token_by_symbol}. Then, for  each candidate address, we query BscScan/Etherscan for transactions within a one-hour window ($\pm30$ minutes around the pump announcement). If only one token shows activity, we label it as the target, while if multiple tokens do, we conservatively discard the event, which occurred 7 times.

\mypara{Source code extraction.}
Our pipeline identified 2,121 distinct target tokens and 3,677 operations in total. 
The number of operations exceeds the number of target tokens because organizers target the same token across multiple operations on different days.
We enriched the dataset with the target token source codes by querying the Etherscan and BscScan APIs. 
These blockchain explorers publish verified source code only when the submitted contract compiles to bytecode that matches the deployed on-chain code. 
We recovered source codes for 1,377 tokens (64.9\% of the dataset). 

\mypara{Trading data collection.}
To reconstruct the economic dynamics of each operation, we collect every on-chain event emitted by the target liquidity pools in our dataset, from pool creation through January 2025.
Every target liquidity pool in our dataset is deployed on Uniswap V2 or one of its forks (\eg PancakeSwap on BSC), which emit four event types relevant to our work: \textit{PairCreated}, when a liquidity pool is initialized, \textit{Mint} and \textit{Burn}, when liquidity is added or removed by liquidity providers, and \textit{Swap}, each time a trade is executed within a liquidity pool.
To guarantee complete historical coverage, we operate two dedicated Geth~\cite{geth} nodes, one for BSC and one for Ethereum, and query them with Web3.py~\cite{web3_py}, retrieving all events associated with each target liquidity pool. 
We collect 2,121 PairCreated, 166,847 Mint, 30,054 Burn, and 4,746,720 Swap events.


\mypara{Pump window definition.}
The collected on-chain events cover the full lifetime of each target liquidity pool. However, only a subset of these events occurs during an operation. We therefore define an operation-specific time window to determine which events should be attributed to each operation.
The Telegram announcement timestamp provides a natural start time, but the end of an operation has no comparable off-chain marker. Prior work either adopts fixed, manually chosen thresholds~\cite{xu2019anatomy} or relies on manual inspection~\cite{la2023doge}. The former does not account for variation in operation duration, whereas the latter does not scale to our dataset. We therefore infer the end of each operation from trading activity, following the volume-based approach of prior work~\cite{kamps2018moon}.
For each operation, we partition the \textit{Swap} events of the target liquidity pool into one-hour bins beginning at the announcement time and compute the hourly trading volume of the pool's base asset (\eg BNB). We define the baseline volume as the mean hourly volume during the preceding week. For pools created less than one week before the announcement, we compute the baseline over the interval between pool creation and the announcement. 
We define the end of the operation as the first post-announcement bin whose trading volume returns to the baseline level. 

        

\section{Trading Activity Analysis}
\label{sec:anatomy_pump_and_dump}

Analyzing trading activity during the operation windows, we observe 528,866 swap transactions executed by 171,755 unique addresses,  totaling \$74.1M in buy volume and \$61.2M in sell volume. 
While buy and sell volumes are relatively balanced, the distribution of trading activity across addresses is highly asymmetric. 
We find 159,831 addresses execute at least one buy transaction, whereas only 73,693 execute at least one sell transaction.

\begin{figure}
    \centering
    \includegraphics[scale=0.27]{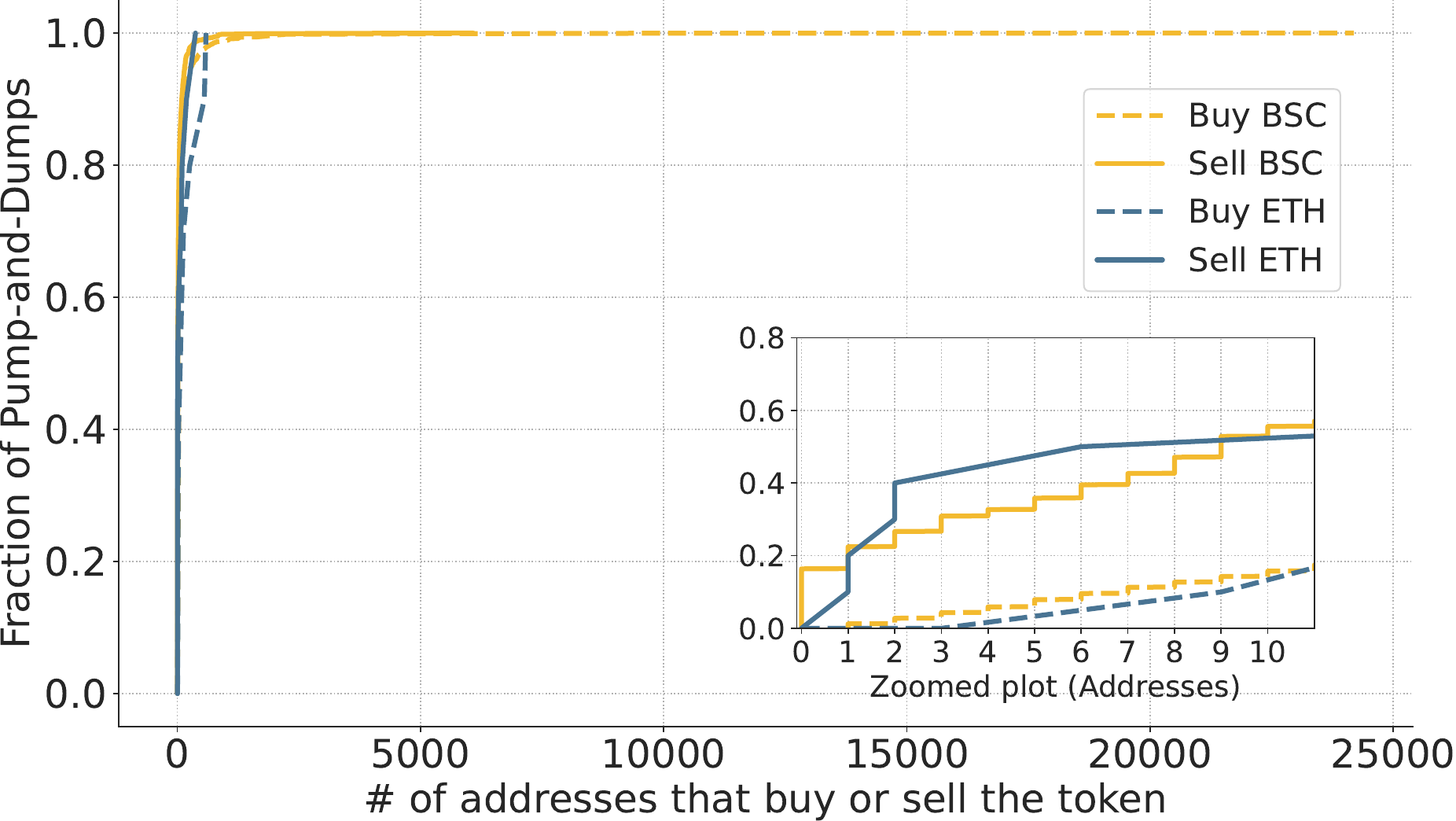}
    \caption{CDF of the number of addresses that buy (dashed) and sell (solid) per operation.}
    \label{fig:cdf_buy_sell}
\end{figure}

To understand the reasons behind the disparity between the number of buy and sell operations, we analyze the number of unique addresses executing each type of transaction within the operation windows. 
Figure~\ref{fig:cdf_buy_sell} shows the cumulative distribution function (CDF) of the number of unique addresses executing buy and sell transactions during each operation window.  Dashed lines represent buy operations, while solid lines denote sell operations.
While many operations attract hundreds of buyers, only a small number of addresses sell.
On BSC, 35.8\% of operations involve five or fewer sellers, compared with only 7.9\% involving five or fewer buyers. 
A similar pattern is observed on Ethereum, where the corresponding proportions are 45.5\% and 9.1\%.

To investigate the cause of this asymmetry, we manually inspect a sample of these events by combining on-chain transaction analysis with the Telegram messages used to coordinate each operation. This joint analysis shows that, in most cases, operations promoted on Telegram as \pd{} actually target honeypot tokens that restrict token sales.
To show concretely how such an operation is carried out, the following section describes a \bt{} in full detail.

\subsection{Dissecting a \bt{} scheme}
\label{sec:first_case_study}

We report a detailed case study of a \bt{} targeting the VAULT~\footnote{0x87A32d9e4DE119261D128018eF9710D87C393B1C} token on BSC, orchestrated by the \textit{CRYPTO PUMP GROUP ALEX WALKER} Telegram channel on October 11, 2022.

\mypara{Initial Funding and Token Creation.}
The analysis begins with the externally owned account~\footnote{0x87d83ac841e2d5672109b988218ce758ddda6d53} (EOA) that created the VAULT token. 
Inspecting its transaction history, we find that all 76 transactions are related to the \bt{}. 
The address's first transaction is an incoming transfer of 50 BNB, received a few minutes prior to the token creation (09:32 UTC). 
Three minutes later (09:35 UTC), the creator performs four transactions to distribute 35 BNB across four addresses with no prior transaction history: three receive 10 BNB each, and the fourth receives 5 BNB. 
At 09:45 UTC, the EOA creates the target token and its associated liquidity pool. 
Two minutes later (09:47 UTC), it provides the initial liquidity by depositing 10 BNB and 120,000 VAULT into the pool.

\mypara{Address Allowlisting.}
Five minutes later (09:52 UTC), the token creator performs four distinct transactions invoking the \texttt{ExcludeFromAll} function of the VAULT smart contract, each time with one of the previously funded addresses as its argument. 
Inspecting the contract's source code, we find that this function modifies the token's transfer behavior. 
Specifically, an address that purchases VAULT tokens receives them only if \texttt{ExcludeFromAll} has previously been invoked for that address; otherwise, the buyer receives nothing, with the invested BNB locked in the liquidity pool.
The four pre-funded addresses, together with the token creator, are thus the only ones able to trade freely, while every other buyer loses their funds.
These four addresses are the \textit{colluding addresses}.

\mypara{Wash trading.}
Almost 50 minutes before the token is announced in the Telegram channel (13:13 UTC), the token creator executes 30 swaps, alternating between buying and selling small amounts, ranging from 0.1 to 0.25 BNB.
This pattern is consistent with wash trading~\cite{la2023game}, a manipulative practice used to artificially inflate trading activity.
The aim is to make the token appear actively traded and sellable.

\mypara{Target Release and Pump Phase.}
The channel administrator announces VAULT as the target at 14:00 UTC.
Immediately after, 33 distinct addresses buy the token, with individual purchases ranging from a few dollars to nearly \$300, for a total of 5.74 BNB (approximately \$1,589).
Because none of these addresses is allowlisted, every purchase transfers zero VAULT tokens, and the invested funds are irreversibly lost.

\mypara{Simulated \pd.}
After the first wave of buy operations (14:05 UTC), the colluding addresses execute large buys of 1 to 3 BNB each, which drive the price spike.
Roughly 15 minutes later (14:21 UTC), they sell their entire holdings, causing a sudden price collapse that makes the operation resemble a conventional \pd{} scheme.

\mypara{Promotion and Message Removal.}  
Shortly after the price collapse, the channel administrator posts a chart advertising a 2,000\% increase (14:47 UTC), presenting the operation as a success (Fig.~\ref{fig:pump_successfull_message} in the Appendix).
Moreover, as illustrated in Appendix in Fig.~\ref{fig:telegram_message_obfuscated}, the administrator deletes the target token's address from the announcement message, justifying the removal as a way to prevent latecomers from buying the token after the operation has ended and incurring losses.
In practice, this action obscures the deceptive nature of the operation by preventing later subscribers from verifying how it ended.
As a result, they must rely on the administrator's charts, which portray it as a successful \pd{} scheme.

\mypara{Funds Collection and Redistribution.} 
Approximately four hours after the operation began (17:59 UTC), the token creator removes the remaining liquidity from the pool to extract all the profits generated.
These funds are then transferred to a different address~\footnote{0x107d3abc61d51f9B6BD80BE959c22DA62eE08872}, which uses them to deploy another token embedding the same \texttt{ExcludeFromAll} honeypot logic and to fund its colluding addresses.
This second token is the target of a \bt{} operation orchestrated by the same Telegram channel and following the same phases, suggesting that the profits of one operation finance the next and that both are under common control.

\section{Detection of \bt{} operations}
\label{sec:detection_honeypot_tokens}
To determine whether \bt{} is an isolated case or a common practice, we develop \tool{}, a tool that emulates token purchases and sales at the blockchain state in which each \pd{} operation occurred. We use it to quantify the extent of \bt{} operations in our dataset.

\subsection{\tool{}}
\label{subsec:tool}

To assess the presence of honeypot tokens in our dataset, we develop and open-source \tool{}~\cite{additional_material}, a tool that emulates token purchases and sales through a liquidity pool and monitors two indicators of honeypot behavior.

The first indicator is whether a transaction reverts. When a swap transaction reverts, its execution fails and any changes made to the blockchain state during execution are discarded, effectively leaving the state as if the swap had not occurred. Sell-side reverts are particularly relevant in the context of honeypots: a trader may successfully purchase a token but later be unable to sell it because the sell transaction systematically reverts.
The second indicator is a significant deviation between the expected and actual amount of assets received. Before each swap, \tool{} estimates the expected output based on the current liquidity pool state and compares it with the actual output. A discrepancy between the expected and actual output may indicate a transfer fee, while a discrepancy approaching the full expected amount may indicate honeypot-like behavior.

We now describe the execution pipeline of \tool{} in detail.

\begin{enumerate}
\item \textbf{Initialization.} 
\tool{} uses Anvil~\cite{anvil}, a local Ethereum development node, to fork the blockchain state at the target block height. Then, it can either create a new externally owned account (EOA) or use a user-specified address to determine whether trading restrictions apply to that address. The address is funded with $\phi$ units of the native cryptocurrency (\eg ETH or BNB).

\item \textbf{Buy Output Calculation.}
\tool{} queries the DEX router's \texttt{getAmountsOut} function with $\phi$ units of the native cryptocurrency as input. Based on the current liquidity pool reserves and the constant-product formula (see Sec.~\ref{sec:background_amm}), the function computes the amount of target tokens expected from the buy operation. We denote this value as $\beta_{exp}$, the expected output of the buy transaction.

\item \textbf{Buy Swap Execution and Fee Estimation.}
\tool{} executes a transaction that calls the DEX router's \texttt{swapExactETHForTokens} function to exchange $\phi$ units of the native cryptocurrency for the target token. After a successful swap, \tool{} queries the target token's \texttt{balanceOf} function to determine the actual amount of tokens received, denoted as $\beta_{act}$. By comparing $\beta_{act}$ with the expected output $\beta_{exp}$, \tool{} estimates the buy fee as $(\beta_{exp}-\beta_{act})/\beta_{exp}$.

\item \textbf{Allowance Approval.}
\label{tool:stepfour}
Before executing the sell swap, \tool{} calls the target token's \texttt{approve} function to grant the DEX router sufficient allowance to spend the acquired tokens. Without this allowance, the router cannot transfer the tokens on behalf of the account, and the subsequent sell operation cannot be performed. Token contracts may include logic that deliberately prevents the required allowance from being granted, causing the approval transaction to revert. In this case, \tool{} records an \emph{approval reversion} and stops the execution.

\item \textbf{Block Advancement.}
After the approval transaction, \tool{} can introduce a delay of $\tau$ seconds before executing the sell operation. The delay is translated into the corresponding number of blocks, which are then advanced on the local fork. This allows \tool{} to detect anti-simulation mechanisms that permit selling only in the same block as the purchase or within a few subsequent blocks.  Without this delay, an immediate sell may succeed and incorrectly suggest that the token is freely sellable. 

\item \textbf{Sell Output Calculation.}
Before performing the sell transaction, \tool{} queries the DEX router's \texttt{getAmountsOut} function a second time. In this call, the actual token balance $\beta_{act}$ is used as input to estimate the expected amount of native cryptocurrency $\sigma_{exp}$.

\item \textbf{Sell Swap Execution and Fee Estimation.}
\tool{} attempts to sell the entire target token balance $\beta_{act}$ for the native cryptocurrency by calling the DEX router's swap function\footnote{\texttt{swapExactTokensForETHSupportingFeeOnTransferTokens}}. If the transaction succeeds, \tool{} records the actual amount of native cryptocurrency received, denoted as $\sigma_{act}$, and estimates the sell fee as $(\sigma_{exp}-\sigma_{act})/\sigma_{exp}$. If the sell transaction reverts, \tool{} records a \emph{sell reversion}.
\end{enumerate}


\subsection{Honeypot token detection}
\label{subsec:experimental-setup}
\subsubsection{Experimental setup}
In this section we leverage \tool{} to detect honeypot tokens in our dataset.
Our goal is to determine tokens behaving as honeypot in our dataset at the time of the \pd{} operation.
Thus, we test each token in the dataset initializing the tool to fork the blockchain at the first block whose mining timestamp is immediately after the administrator released the target coin in the channel.
Using the blockchain height at the time the \pd{} operation starts on Telegram, we ensure that the emulation reflects the exact conditions under which real participants interacted with the token.
Then, to emulate a realistic purchase while avoiding artifacts introduced by extremely small trades, we set the trade amount, $\phi$, relative to the pool liquidity at the time of the operation rather than using a fixed absolute value. Specifically, we use 0.1\% of the pool liquidity, with a minimum trade value of \$10 for low-liquidity pools. This choice keeps $\phi$ well below common anti-whale thresholds~\cite{cernera2025blockchain}, which are typically expressed as a percentage of the total supply or pool reserves, reducing the likelihood of misclassifying legitimate anti-whale protections as honeypot behavior while ensuring that the simulated trade remains representative of normal user activity.
We first ran the tool with $\tau = 0$ to identify tokens whose sell transactions revert regardless of the elapsed time, establishing a baseline. We then repeated the analysis with $\tau$ set to 1, 5, 10, 30 seconds, 1 and 10 minutes.
Repeating the analysis multiple times is important as it is possible that token smart contracts implement anti-simulation mechanisms~\cite{cernera2025blockchain}.

 


\begin{figure}   
    \centering
    \includegraphics[scale=0.27]{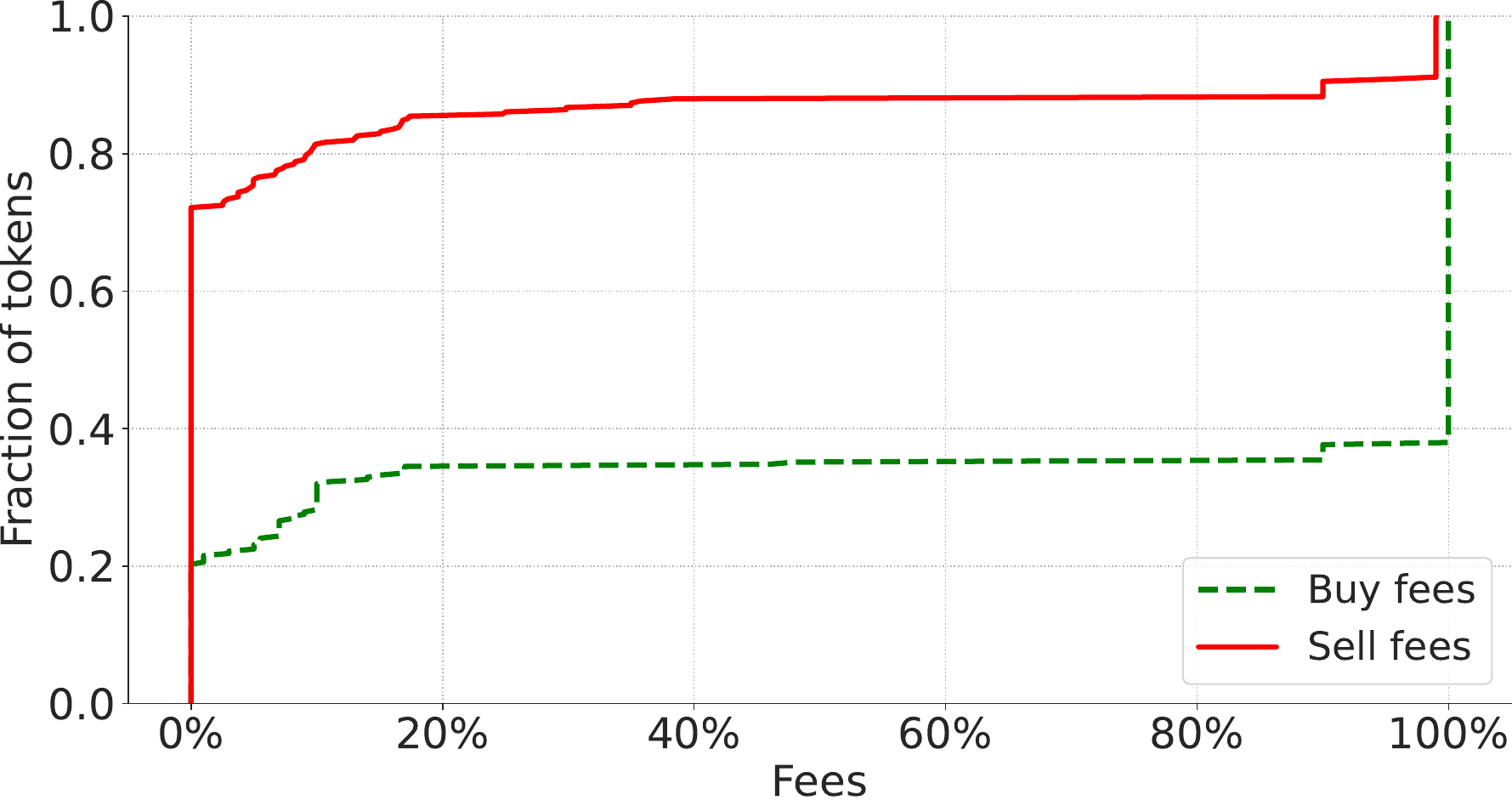}
    \caption{Distribution of fees found in tokens.}
    \label{fig:cdf_buy_sell_taxes}
\end{figure}

 \subsubsection{Detection Results and Honeypot Mechanisms}
\label{subsec:tool_results}

\mypara{Reversion-based detection.}
With $\tau = 0$, \tool{} identifies 1,095 out of 2,121 tokens for
which the simulated trading cycle cannot be completed. Specifically,
the sell transaction reverts for 983 tokens, while the approval
transaction reverts for 112 tokens. Increasing $\tau$ reveals
additional tokens whose sell restrictions become effective only after
a delay. Relative to the $\tau = 0$ configuration, the number of
detected tokens increases cumulatively by 62, 187, and 191 when
$\tau$ is set to 1, 5, and 10 seconds, respectively. Thus, with
$\tau = 10$ seconds, \tool{} detects 191 additional honeypot tokens
compared with the immediate-sell configuration. We observe no further
changes for larger values of $\tau$. Timed sell reversions therefore
bring the total number of tokens detected through transaction
reversions to 1,286.

\mypara{Fee-based detection.}
For tokens whose approval and sell transactions succeed for every
tested value of $\tau$, we analyze the difference between the expected
and actual outputs of the simulated trades.
Figure~\ref{fig:cdf_buy_sell_taxes} reports the cumulative
distributions of the estimated buy- and sell-side fees. The
distributions exhibit a clear separation between tokens imposing no or
moderate fees and tokens imposing fees close to 100\%. In particular,
we observe a gap between the moderate-fee region and the concentration
of tokens with fees at or near 100\%, suggesting the presence of two
distinct behavioral groups.

Based on this separation, we conservatively classify a token as an
economically prohibitive honeypot when its simulated buy- or sell-side
fee exceeds 90\%. This threshold captures cases in which a transaction
technically succeeds but produces an outcome economically equivalent
to losing the invested funds. At the same time, the high threshold
reduces the risk of classifying tokens with substantial but potentially
legitimate transfer fees as honeypots. This criterion identifies 206
additional tokens: 180 with a high buy-side fee and 26 with a high
sell-side fee.

\mypara{Hierarchical classification.}
We assign each token-operation pair to exactly one of five mutually exclusive categories: approval reversion, immediate sell reversion, timed sell reversion, high buy-side fee, or high sell-side fee. We first classify cases in which the approval transaction reverts. Among the remaining cases, we classify as immediate sell reversions those that revert with $\tau = 0$, and as timed sell reversions those that succeed with $\tau = 0$ but revert for a positive value of $\tau$.
Fee-based categories are considered only when both the approval and sell transactions succeed for every tested value of $\tau$.

We execute \tool{} independently for every operation at its
corresponding historical blockchain state, including operations that target the same token multiple times.
Applying this classification, we identify 2,629 \bt{} operations,
corresponding to 71.5\% of all operations in our dataset. These operations involve 1,492 distinct tokens, or 70.3\% of all targeted tokens. Of these, 1,286 tokens are detected through transaction reversions and 206 through economically prohibitive fees. The remaining 1,048 operations involve 629 tokens for which \tool{} identifies no honeypot behavior under the tested conditions.
Table~\ref{tab:table_honeypot_mechanism} reports the complete breakdown.

To further investigate these findings and understand how the identified behaviors are implemented in practice, we analyze the dataset of verified smart contracts described in Sec.~\ref{sec:dataset}. In Appendix~\ref{appendix:implementation_anal}, we describe the code-level mechanisms underlying the runtime behaviors detected by \tool{}. We also publicly release the complete source code of a representative real-world token implementation for each identified honeypot mechanism~\cite{additional_material}.

\subsubsection{Validation}
\label{subsec:validation}

To evaluate the behavior of our detection methodology on reference tokens outside the Telegram corpus, we use two datasets: a publicly available dataset of tokens labeled as honeypots and a curated reference set of tokens that can be assumed to be benign.

To the best of our knowledge, the only publicly available dataset of
labeled honeypot tokens is the one introduced by Huynh
et al.~\cite{huynh2025programming}. The dataset contains 1,859 distinct
tokens, each associated with a distinct liquidity pool. Since the
liquidity of these pools has subsequently been drained, evaluating the
tokens at the current blockchain state would not reproduce a tradable
market state. We therefore fork the blockchain immediately after the
block in which liquidity was first added to each pool, corresponding to
the earliest state in which the token could be traded through that pool.

We evaluate both reference datasets using the same \tool{} configuration
and classification criteria as in the main analysis, with $\tau=10$
seconds, since larger values yielded no additional timed sell reversions
(Sec.~\ref{subsec:tool_results}). Using this configuration, we
evaluate the 1,859 tokens in the Huynh et al.~dataset at the historical
state immediately following their initial liquidity provision. The
simulated buy-approve-sell sequence satisfies our honeypot
classification criteria for 1,676 tokens (90.2\%). This result measures
agreement between our behavioral criteria, evaluated at the selected
historical state, and the external honeypot labels.

We then evaluate \tool{} on tokens that can reasonably be assumed to be
benign. Since we could not identify a suitable publicly available
dataset of benign tokens, we collect 967 tokens from PancakeSwap's
official token list~\cite{pancakeswap_list}. This list contains tokens
displayed in the PancakeSwap web interface whose inclusion is reviewed
by the PancakeSwap core team according to the platform
documentation~\cite{pancakeswap_list_faq}. We run \tool{} on these
tokens in July 2026. The tool correctly classifies all 967 tokens as
non-honeypot.

Across the two reference datasets, we observe no positive classification that contradicts the provided labels: \tool{} identifies honeypot-like behavior only among tokens independently labeled as honeypots, and raises no alerts on the benign-token reference set. 
We analyze the 183 labeled honeypot tokens not detected by
\tool{} in Sec.~\ref{sec:limitations}. 

\begin{table}
    \caption{\bt{} and \pd{} operations detected and breakdown of honeypot implementations.}
    \label{tab:table_honeypot_mechanism}
    \centering
    \small
    \begin{tabular}{l|rr}
        \toprule
        \textbf{Category}   & \textbf{\# of operations} & \textbf{\# of tokens} \\
        \midrule
        \textbf{\bt{}} &   \textbf{2,629 (71.5\%)} & \textbf{1,492 (70.3\%)} \\
        \addlinespace[0.2em]
        \quad  -- Sell reversion  & 1,902 (72.3\%) & 983 (65.9\%) \\
        \quad  -- Timed sell reversion  & 273 (10.4\%) & 191 (12.8\%) \\
        \quad  -- Approval reversion  & 120 (4.6\%) & 112 (7.5\%) \\
        \quad  -- High buy fee ($>90\%$)  & 288 (11.0\%) & 180 (12.1\%) \\
        \quad -- High sell fee ($>90\%$)  & 46 (1.7\%) & 26 (1.7\%) \\
        \midrule
        \textbf{\pd{}} &   \textbf{1,048 (28.5\%)} & \textbf{629 (29.7\%)} \\
        \bottomrule
    \end{tabular}
\end{table}

\section{Characterizing \bt{} Operations}
\label{subsec:conventional-vs-honeypot}
In this section, we characterize \bt{} operations by contextualizing our findings through a comparison with the \pd{} operations in our dataset.

\begin{figure}   
    \centering
    \includegraphics[scale=0.24]{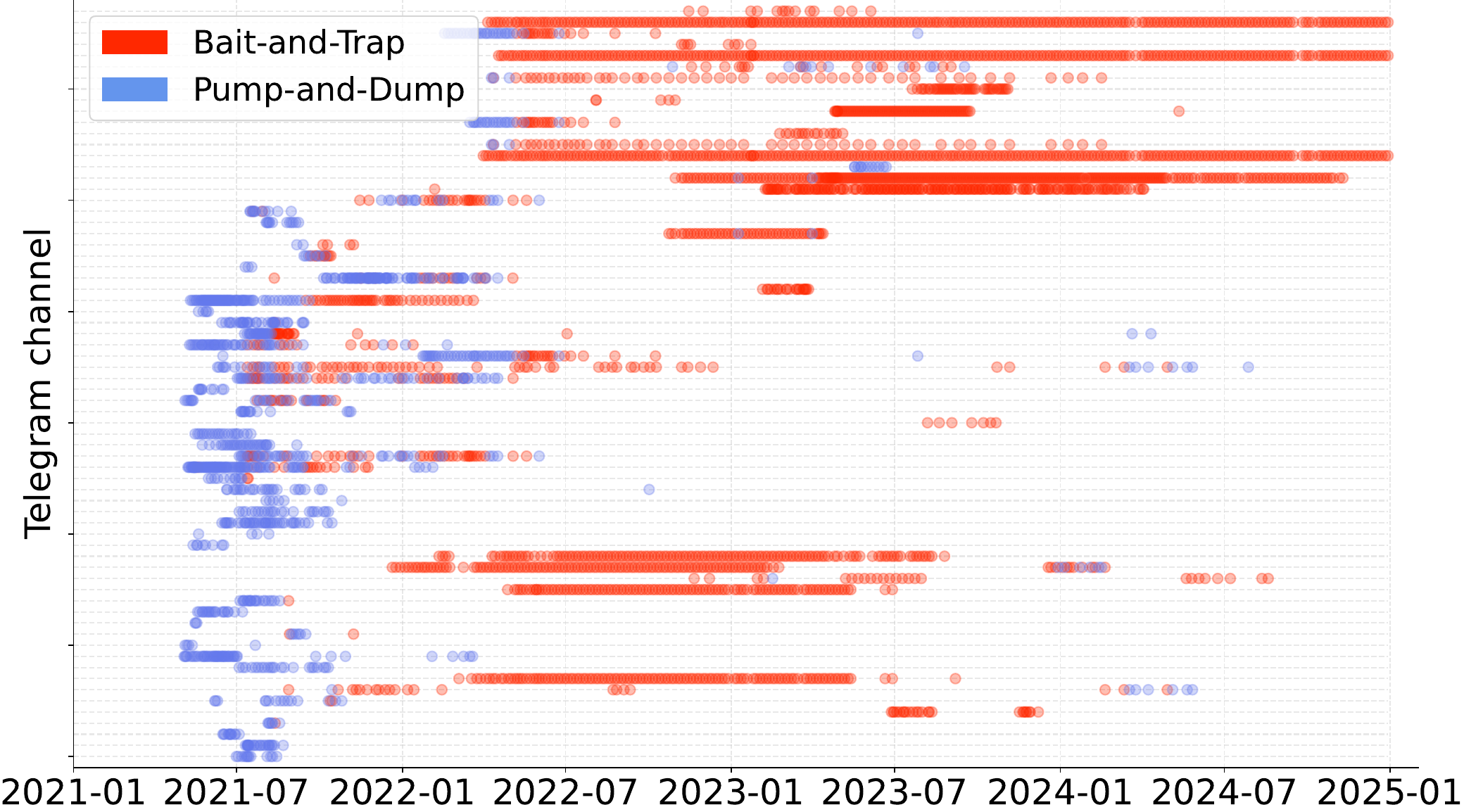}
    \caption{\bt{} and \pd{} events performed by Telegram channels over time.}
    \label{fig:pump_over_time}
\end{figure}

\mypara{\bt{} has become the dominant operation type over time.}
Figure~\ref{fig:pump_over_time} shows the \pd{} operations carried out by Telegram channels over time. Each line corresponds to a Telegram channel, and each point represents a single operation organized by that channel\footnote{Channels with fewer than three operations are excluded to improve readability.}. Blue points denote \pd{} operations, while red points denote \bt{} operations.
Two observations emerge from the figure. First, channels conducting \bt{} operations are substantially more prolific than those organizing \pd{} schemes, with the most active channel carrying out 296 operations. Second, the figure reveals a clear temporal shift in the dominant fraud mechanism. \pd{} operations (blue points) are heavily concentrated in the early part of the observation period, with 94.5\% occurring before June 2022. From mid-2022 onward, the landscape is instead dominated almost entirely by \bt{} operations (red points).
This trend suggests that \bt{} schemes have progressively replaced \pd{} operations in our dataset. One possible explanation is that early Telegram channels initially replicated the \pd{} model previously observed on CEXs, whereas newer channels increasingly adopted the more profitable \bt{} strategy, an aspect that we analyze in Sec.~\ref{subsec:profits}.

\begin{figure}
\centering
\includegraphics[scale=0.24]{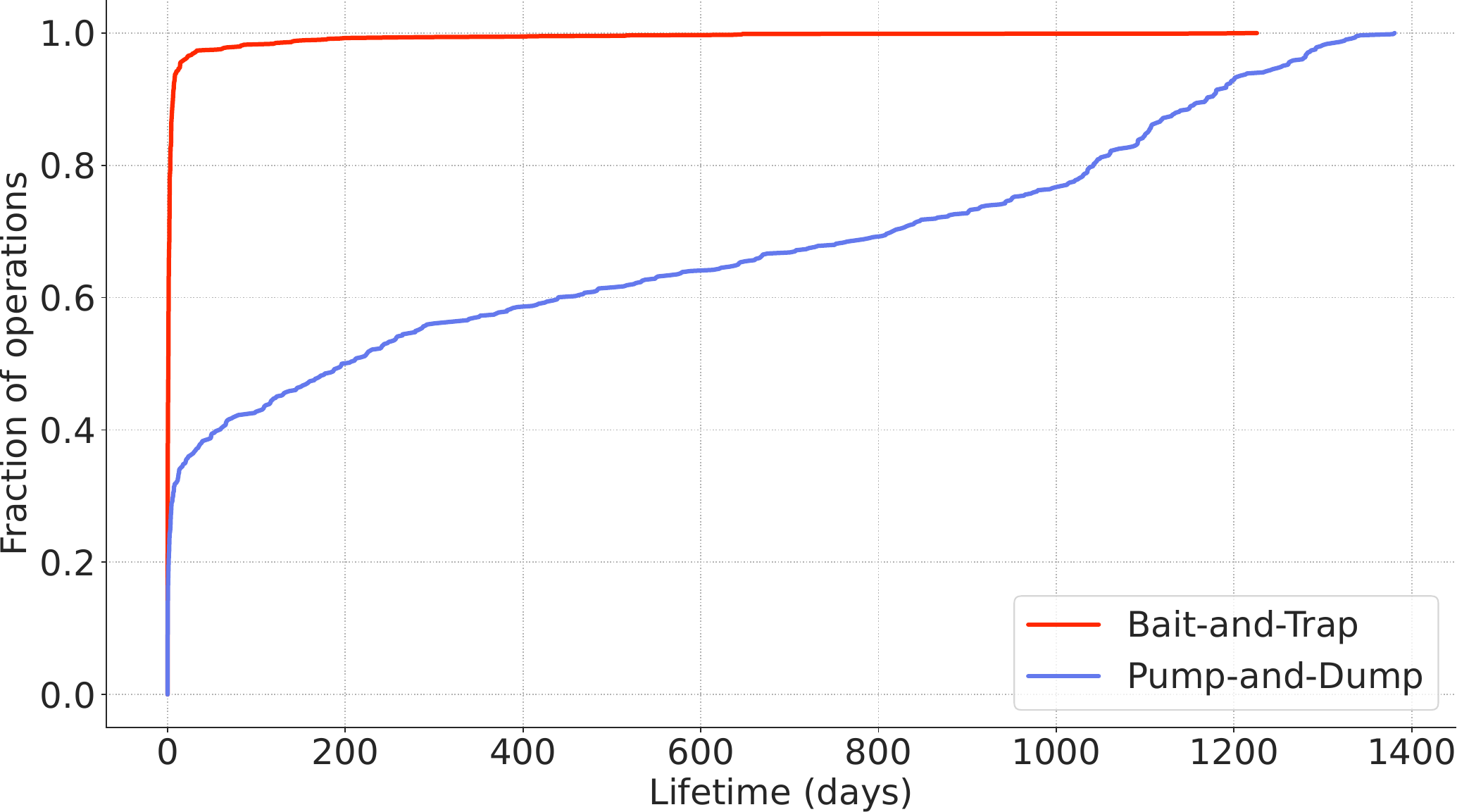}
\caption{Lifetime of tokens targeted by \bt{} and \pd{} operations.}
\label{fig:cdf_distance_lifetime}
\end{figure}

\mypara{\bt{} targets disposable tokens created specifically for the operation.}
We analyze the market lifetime of targeted tokens to measure how long they remain actively traded. 
We define this metric as the time elapsed between the creation of the liquidity pool targeted by the operation and the last recorded market activity of the token (\ie the last emitted Mint, Burn, or Swap event).
Figure~\ref{fig:cdf_distance_lifetime} shows the lifetime distribution of tokens targeted by \bt{} operations (red) and \pd{} operations (blue). The figure reveals a substantial difference between the two categories: \bt{} operations overwhelmingly target short-lived tokens, with a median lifetime of only one day. In particular, 79.6\% of targeted tokens have a lifetime of less than 72 hours and 91.8\% of less than one week, indicating that the token is created, exploited, and abandoned within days.
In contrast, tokens targeted by \pd{} operations exhibit a median lifetime of 211 days, suggesting that they are established assets with ongoing trading activity.
We further investigate whether the target tokens of \bt{} operations are created specifically for the manipulation by measuring the time elapsed between token deployment and the corresponding \pd{} event. Figure~\ref{fig:cdf_distance_token_creation_pump}, in the Appendix, reports the cumulative distribution of this interval. We find that honeypot tokens are typically deployed shortly before the manipulation: 91.1\% are created within 48 hours of the corresponding event.
This pattern suggests that tokens targeted by \bt{} operations are disposable assets designed specifically for the manipulation. 


\mypara{Apparently unrelated \bt{} operations are connected.}
As shown in the case study (Sec.~\ref{sec:first_case_study}), profits extracted by the token creator are transferred to fresh addresses, which are subsequently used to conduct additional \bt{} operations. 
We therefore perform a flow-tracing analysis~\cite{wu2021detecting,meiklejohn2013fistful} starting from the token creator of each \bt{} operation to identify relationships between distinct operations. 
We trace all addresses that receive transfers of ETH, BNB, and stablecoins from the token creator after the operation has been completed. If any of these recipient addresses subsequently act as the token creator of another \bt{} operation, we consider the two operations grouped into the same money-flow cluster.

To prevent the tracing process from expanding indefinitely and introducing unrelated addresses, we apply two constraints. First, we limit the tracing depth to two hops: if no address involved in a \bt{} operation is found within two transfer hops, the trace terminates. Second, we build a known service dataset (Appendix~\ref{subsec:known-service-dataset}) to stop tracing whenever the flow reaches an address belonging to a known service (\eg custodial wallets). Such services aggregate and redistribute funds from many unrelated users, making them taint boundaries beyond which fund flows are no longer informative.
The analysis identified 70 money-flow clusters containing 1,661 operations, approximately 63\% of all \bt{} operations.
The clusters vary substantially in size, with the top four clusters containing more than 100 connected operations each.

\begin{figure}
\centering
\includegraphics[scale=0.12]{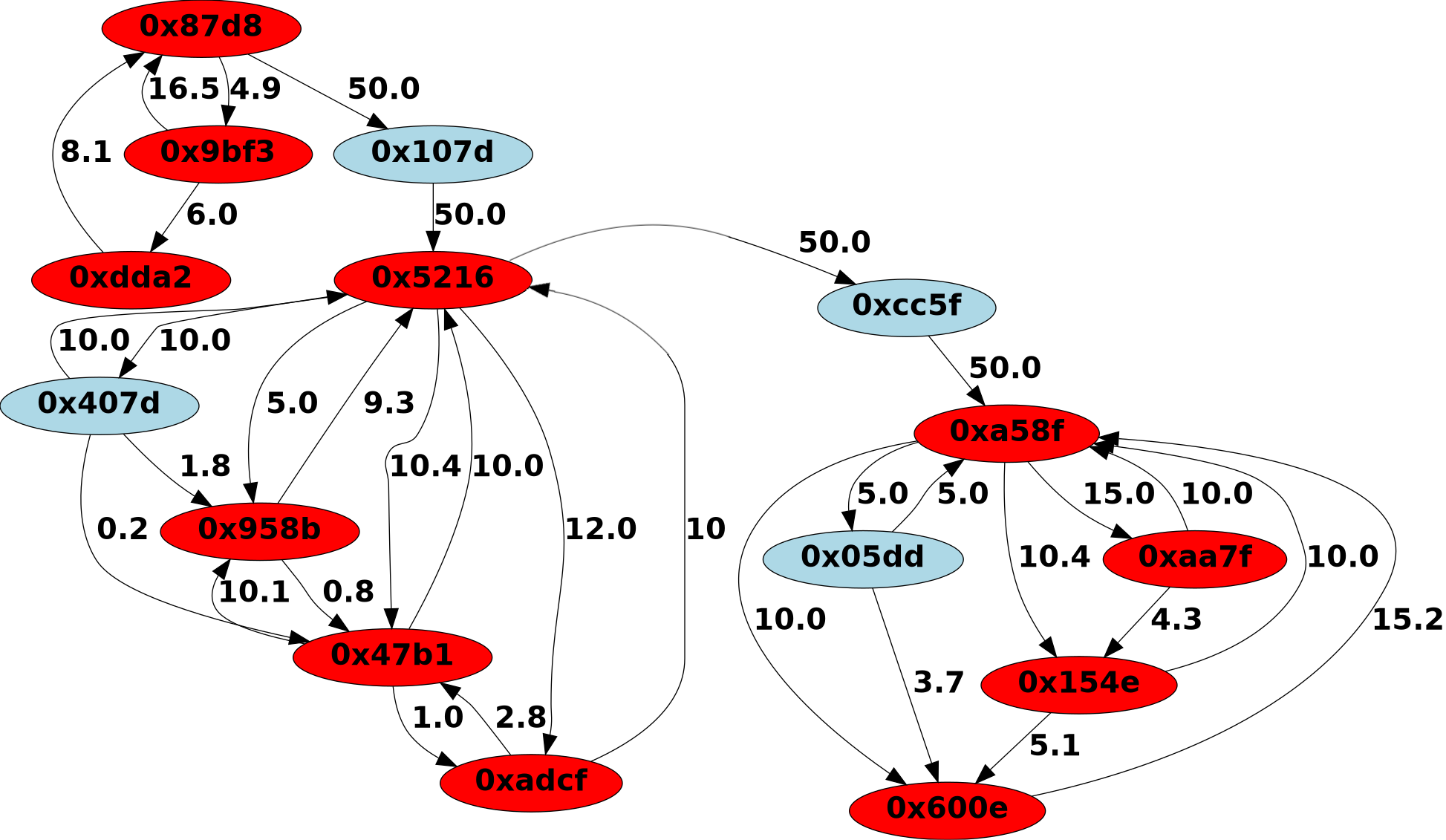}
\caption{A subset of the addresses reached from node \textit{0x87d8} obtained through fund-flow analysis.
}
\label{fig:small_case_study}
\end{figure}

Tracing the flow of funds from the organizer's address in the case study (Sec.~\ref{sec:first_case_study}), our analysis identifies 27 addresses that create new tokens later targeted by \bt{} operations in our dataset. These operations follow the same pattern described in detail in the case study.
Figure~\ref{fig:small_case_study} illustrates the resulting graph for the first 11 operations in this chain. In the figure, each node represents an address. Red nodes correspond to addresses that are used to perform \bt{} operations, while blue nodes represent addresses that, to the best of our knowledge, act solely as intermediaries for fund transfers. Edges represent transactions between addresses and are annotated with the amount of BNB transferred.
These connected manipulations occurred between 2022-10-11 and 2022-11-25, with one operation executed approximately every two days. The schemes were promoted alternately across two Telegram channels: 14 operations were advertised on the \textit{"CRYPTO PUMP GROUP ALEX WALKER"} channel, while the remaining 13 appeared on the \textit{"Crypto Pumps Crypto Community"} channel.
The total profit for the organizer of these operations was \$80,069 over 45 days.

\section{Profits}
\label{subsec:profits}
In this section, we estimate the gains and losses associated with \pd{} and
\bt{} operations. As shown in the case study
(Sec.~\ref{sec:first_case_study}), colluding addresses generate a
substantial fraction of the trading activity observed during an operation. We
therefore first identify these addresses, both to quantify their prevalence and
to account for their activity when estimating the profits of \bt{}
operations.

%
\subsection{Detection of colluding addresses}
\label{subsec:detection_colluding_addresses}

To identify colluding addresses, we perform two complementary analyses.
In the first, we examine the smart contracts of honeypot tokens looking for hard-coded lists of addresses that are not affected by the honeypot restriction logic. 
In particular, we use regular expressions to extract all hard-coded addresses present in the source code, obtaining 1,349 addresses. To remove noise related to addresses unrelated to the \bt{} operation (\eg addresses that automatically receive a portion of each transaction as a marketing or development fee), we retain only those that perform at least one trading operation during the operation window. After this filtering step, we obtain a set of 1,316 addresses.

In the second analysis, we account for addresses that can be dynamically added to the allowlist after token deployment by invoking specific smart contract functions. These functions are used to update the set of colluding addresses. To identify addresses added through this mechanism, we first collect all transactions sent to honeypot tokens that invoke smart contract functions, obtaining a total of 129,090 transactions. For each transaction, we extract the function selector from the input data field. The selector corresponds to the first 4 bytes of the Keccak-256 hash of the invoked function signature and identifies the called function.
To decode the function selectors, we follow the approach proposed by Wu et al.~\cite{wu2025hunting} and leverage the APIs provided by the Ethereum Signature Database~\cite{ethereum_signature_database}, which maintains mappings between function selectors and their corresponding human-readable function signatures. Using this approach, we recovered 633 unique function signatures.
We then manually inspected the previously recovered smart contract source code to locate occurrences of the identified function signatures and determine which were explicitly designed for allowlist management (\eg \texttt{addToWhitelist(address)}, \texttt{whitelistBatch(address[] addrList)}). Through this process, we identified 26 functions used to dynamically modify allowlist contents. To ensure reproducibility, the complete list of identified signatures is available in our public repository~\cite{additional_material}.
Once the relevant functions were identified, we parsed the full bytecode of the corresponding transactions. By applying the recovered function signatures as an interface, we interpreted the input parameters according to their specific data types (\eg \texttt{address} or \texttt{address[]}), allowing us to extract the actual addresses passed as arguments. Using this approach, we recovered a total of 2,742 unique colluding addresses.
Combining both analyses, we discovered 4,039 unique colluding addresses across 1,100 tokens.  As a final validation, for each operation in which an address is identified as colluding, we simulate buy and sell transactions using that address instead of a freshly generated one, confirming that, unlike the fresh address, it can both buy and sell freely.

\mypara{The impact of colluding addresses.}
Across all \bt{} operations, we observe 38,080 unique addresses interacting with the targeted liquidity pools, of which only 4,039 (10.6\%) are colluding.
However, their impact becomes clear when examining trading volumes. 
Colluding addresses generate a total buy volume of \$27.3M and a sell volume of \$16.6M, compared to just \$10.3M in buy volume and \$1.3M in sell volume for non-colluding addresses.
They account for approximately 72.5\% of the total buy volume and 92.6\% of the total sell volume, confirming that the trading activity observed during \bt{} operations is largely artificial.


\subsection{\bt{} Profits}
To obtain an accurate estimate of the actors participating in the operation, we categorize addresses according to their role within the scheme. We aggregate the address that deployed the token contract and all associated colluding addresses into a single \textit{Token creator} persona. In contrast, the \textit{Investor} group comprises all non-colluding addresses that perform at least one trade within the pool during the operation window.

To compute the profit of each participant, we sum the total amount of base currency received (\eg through liquidity withdrawals or token sales) and subtract the total amount of base currency spent (\eg on liquidity additions or token purchases), as well as transaction fees.
We then convert the resulting net amount to US dollars using the end-of-day exchange rate reported by CoinMarketCap~\cite{CoinMarketCap}, a widely used cryptocurrency market data provider, for the day on which the operation occurred. 
We define the base currency as the asset paired with the target token in the liquidity pool (\eg BNB or a stablecoin in our dataset).
Tab.~\ref{tab:table_gain_honeypot} presents the gains, the losses, and the number of addresses involved in \bt{} operations. 
Focusing on these operations we observe the following insights.

\mypara{Token creators profit in the majority of cases.}
Token creators realize positive returns in 99.3\% of operations, with total gains exceeding \$7M and an average profit of \$2,721 per operation. In the rare cases where token creators are unprofitable, losses are negligible, averaging only \$10 per operation. These losses typically occur when the creator fails to attract sufficient investor participation, incurring only the transaction fees associated with the smart contract deployment, liquidity pool creation, and generation of artificial trading volume.

\mypara{Almost all investors lose money in \bt{} operations.}
In contrast, investors in \bt{}{} operations almost always incur losses. While 13.7\% technically make a profit, the amounts are negligible: the average profit is just \$13 per operation, with a median of \$0.9. These are cases where honeypots are implemented through high taxes, and the price is pushed so high by colluding addresses that early buyers, despite losing most of their tokens to the fee, can obtain  a very small profit.
The majority of investors (86.3\%) incur losses, totaling more than \$7.2M, with an average loss of \$111 per operation.

\begin{table}
    \caption{Overview of the gains (↑) and losses (↓) of participants in \bt{} operations.}
    \label{tab:table_gain_honeypot}
    \centering
    \resizebox{\columnwidth}{!}{
    \begin{tabular}{l|rrr}

        \toprule
         \textbf{Category} & \textbf{Total} & \textbf{Average} & \textbf{Addresses} \\
        \midrule
        Token creator ↑  & \$7,104,516 & \$2,721 & 2,611 (99.3\%) \\
        Token creator ↓ & -\$176 & -\$10 & 18 (0.7\%) \\
        \midrule
        Investors ↑ & \$137,965 & \$13 & 10,356 (13.7\%) \\
        Investors ↓ & -\$7,218,546 & -\$111 & 65,104 (86.3\%) \\
        \bottomrule
    \end{tabular}
    }
\end{table}
\begin{table}
    \caption{Overview of the gains (↑) and losses (↓) of participants in \pd{} operations.}
    \label{tab:table_gain_conventional}
    \centering
    \resizebox{\columnwidth}{!}{
    \begin{tabular}{l|rrr}

        \toprule
         \textbf{Category} & \textbf{Total} & \textbf{Average} & \textbf{Addresses} \\
        \midrule
        Early Investors ↑ & \$3,495,619 & \$481 & 7,263 (60.4\%) \\
        Early Investors ↓ & -\$1,072,042 & -\$225 & 4,755 (39.6\%) \\
        \midrule
        Late Investors ↑ & \$8,789,890 & \$351 & 25,013 (13.4\%) \\
        Late Investors ↓ & -\$22,354,976 & -\$138 & 161,741 (86.6\%) \\
        \bottomrule
    \end{tabular}
    }
\end{table}

\subsection{\pd{} Profits}
For \pd{} operations, we cannot clearly identify which addresses are involved in organizing the manipulation. Therefore, we infer participant roles based solely on the timing of their activity.
Specifically, we define two categories: \textit{Early investors} and \textit{Late investors}. Early investors are addresses that acquire the token up to one week before the \pd{} begins and sell during the \pd{} window. This group may include organizers, as well as regular investors who bought the token in advance.
In contrast, late investors are addresses that purchase the token after the \pd{} has started. This group includes both informed participants who follow Telegram channels and uninformed external investors attracted by the rapid price increase during the pump phase.
Tab.~\ref{tab:table_gain_conventional} presents profits for \pd{} operations and shows the following insights:

\mypara{Investing early does not guarantee a profit in \pd{} operations.} 
The majority of early investors (60.4\%) have a gain in \pd{} operations. However, this success rate is substantially lower than the 99.3\% rate observed for token creators in \bt{} operations. Moreover, the average profit of early investors in \pd{} operations (\$481) is less than a fifth of the average profit earned by token creators in \bt{} (\$2,721).
These findings indicate that executing a \pd{} scheme is significantly less profitable even for the \pd{} organizer.  Moreover, although early investors can gain a competitive advantage by acquiring the target token before the \pd{} operation begins, this strategy remains risky.
Indeed, since the token can be sold freely, other investors may exit sooner
than expected, initiating the dump phase and pushing the price down before
early investors can liquidate their position. Furthermore, if the operation fails to attract sufficient buying interest, the price may not increase enough for early investors to realize a profit.



\mypara{Late Investors incur extremely large losses.} 
Late investors in \pd{} operations experience severe losses, exceeding \$22M. Only 13.4\% achieve positive returns. 
However, when they do profit, their average gain of \$351 is substantially higher than the \$13 observed in \bt{} operations. This highlights a key difference between the two settings: \pd{} operations still allow an opportunity for participants who are fast enough in buying and selling to make a significant profit.

\section{Applications of \tool}
\label{sec:practical-relevance}

\subsection{Historical Token Analysis}
\label{subsec:for_researchers}

When analyzing a past malicious operation, such as a \pd{} operation, it is essential to determine whether the token involved was a honeypot at the time the scheme occurred. 
This is crucial because contract creators can disable trading restrictions, while the liquidity pools used in the scheme are often drained once it ends.

However, we find that existing tools do not support this type of historical analysis. 
We  review academic approaches based on transaction simulation~\cite{gan2023trick,huynh2025serial} and find that none of them publicly release their implementations for use.
Several academic works~\cite{huynh2025programming,mongardini2026midsummer,zhou2025no,10.1145/3639477.3639722} rely on commercial tools, such as Honeypot.is~\cite{honeypotis} and GoPlus Security~\cite{goplus}, to detect honeypot tokens. These tools operate on the current state of a token's contract and liquidity pool, rather than reconstructing their state at a specific point in the past.
\tool{} addresses this limitation by allowing users to analyze tokens at a specific point in the past. Since its implementation and detection logic are released as open source, it is possible to inspect and reproduce the analysis, thereby improving the transparency of the results.

To assess the extent to which the use of commercial tools affects the detection of \bt{} operations, we run them on both our dataset and the validation dataset (Sec.~\ref{subsec:validation}) and compare the results.

\mypara{Honeypot.is.}
Honeypot.is, similarly to our tool, performs transaction simulations to detect honeypots. Given a token address and a target blockchain, it simulates buy and sell transactions and applies additional proprietary checks whose internal logic is not publicly disclosed. The system assigns each token a categorical risk label: low, medium, high, very high, honeypot, or unknown.
When applied to the tokens in our dataset, Honeypot.is identifies 856 tokens as honeypots, compared with 1,492 identified by our tool. Thus, Honeypot.is detects 43\% fewer honeypots than our tool. Among the 636 tokens identified as honeypots by our tool but not by Honeypot.is, 388 (61\%) are classified as unknown, while 221 (34.8\%) are classified as low risk. On the validation dataset, which contains 1,859 tokens, Honeypot.is identifies 1,620 (87.1\%) as honeypots, compared with 90.2\% identified by our tool. 
Manual inspection of a subset of these cases suggests two possible explanations for this discrepancy. First, tokens classified as unknown often cannot be tested because their liquidity pools no longer contain sufficient liquidity to execute the simulated transactions. Second, tokens classified as low risk may no longer exhibit the honeypot behavior that was active at the time of the malicious operation.

\mypara{GoPlus Security.}
GoPlus Security reports whether a token is a honeypot and provides additional risk indicators derived from smart contract analysis that are intended to identify potentially malicious token implementations.
When applied to our dataset, GoPlus Security labels only 138 tokens as honeypots, compared with 1,492 identified by our tool. On the validation dataset, GoPlus identifies 1,107 tokens (59.6\%) as honeypots.
These results show that GoPlus identifies substantially fewer honeypots than our tool on both datasets.

The comparison with both commercial tools confirm that reconstructing past blockchain states is essential for accurate historical analysis, as relying only on present-state simulations would lead to incomplete or misleading results. 


\subsection{Pre-purchase Screening}
\label{subsec:for_investors}

Beyond supporting the analysis of past honeypot schemes, \tool{} can also help investors assess a token before purchasing it. 
In this setting, \tool{} emulates buy and sell transactions against the current blockchain state to determine whether the token exhibits honeypot behavior. This provides investors with a tool to avoid tokens that prevent or severely restrict selling after purchase.

To make this functionality accessible to non-technical users, we provide a web-based interface that allows users to analyze tokens directly from their browser~\cite{web_interface}. Users can specify the amount of native asset used in the simulated purchase ($\phi$ in Sec.~\ref{subsec:tool}), the number of seconds between the simulated buy and sell transactions (computed like the parameter $\tau$ in Sec.~\ref{subsec:tool}), and the address used to perform the simulation. The tool then reports the outcome of the simulated transactions, allowing users to assess whether the token can be sold under the simulated conditions.

To further reduce the effort required from users, \tool{} could be integrated directly into platforms where users discover and interact with tokens, such as DEXs and blockchain explorers. Such an integration could automatically screen a token before a user submits a purchase transaction and present the result alongside other token information. By moving detection to the point of purchase, \tool{} could provide investors with a warning before they commit funds to a potentially malicious token.

A complementary direction is integration within Telegram, which plays a central role in coordinating these schemes. In this environment, a bot could automatically analyze tokens as they are promoted in channels 
and warn subscribers. By shifting detection to the earliest point of exposure, our tool can effectively hinder the attack.

\section{Discussion}
\label{sec:discussion}
\mypara{Why are \pd{} and \bt{} so prevalent on BSC?}
Our observation that BSC is a target of this  activity stems directly from our data collection process. As described in Sec.~\ref{sec:dataset}, we did not target any specific blockchain. Instead, we collected \pd{} schemes coordinated via Telegram channels and subsequently identified the blockchains on which these operations were executed. The concentration of \pd{} activity on BSC, also observed by prior work~\cite{rath2024profiling,li2021cryptocurrency}, can be explained by several factors. First, BSC offers low transaction fees and short confirmation times, enabling rapid, low-cost trading and making the deployment of short-lived, manipulated tokens economically viable. Second, BSC has consistently ranked among the largest blockchain ecosystems by total value locked (TVL) and DEX activity~\cite{yusandika2025onchain}, providing access to substantial liquidity and a large pool of potential victims.

\mypara{Is it possible to prevent honeypot tokens creation?}
The root cause of honeypot tokens is the 
programmability of ERC-20 contracts, which allows operators to embed arbitrary trading restrictions into standard functions. A long-term direction to explore is the development of token standards with stronger behavioral guarantees, for instance, restricting the modifications that can be applied to transfer logic, that would make it structurally impossible to implement honeypot mechanisms.
A step in this direction is ERC20-R~\cite{wang2022erc}, an informally proposed token standard requiring that any transfer violating predefined rules reverts completely.
While such a standard would require broad ecosystem adoption to be effective, the financial harm documented in this paper provides a concrete motivation for research in this direction.

\mypara{Can you link the creation of honeypot tokens directly to the administrators of \pd{} channels?} We cannot establish a definitive link between the administrators of Telegram \pd{} channels and the blockchain addresses responsible for deploying honeypot tokens. However, there is strong evidence pointing to a close connection between them.
First, there is clear temporal coordination between token deployment and \bt{} execution: honeypot tokens are often created just a few hours before the operation begins, making coincidence unlikely and suggesting direct involvement (see Fig.~\ref{fig:cdf_distance_token_creation_pump}). Second, when examining the addresses that deploy these tokens, we observe that several Telegram channels repeatedly target tokens created by the same small set of addresses. This is supported by the high concentration of token creation: only 1,062 addresses deploy all 1,492 honeypot tokens in our dataset, with the top five alone responsible for 238.
Moreover, our flow-tracing analysis (Sec~\ref{subsec:conventional-vs-honeypot}) highlighted links between apparently unrelated operations, further corroborating the hypothesis that channels consistently target tokens created by the same entities.


\section{Limitations}
\label{sec:limitations}
We acknowledge that, despite our efforts to compile a comprehensive keyword list, our search may not capture all channels organizing \pd{} schemes. However, this limitation is inherent to any keyword-based discovery methodology and is shared by prior work~\cite{la2023doge, xu2019anatomy}.  Accordingly, the findings reported throughout the paper should be interpreted as characterizing the present corpus rather than the broader DEX ecosystem, and likely represent conservative estimates of the true scale of the phenomenon.

To better understand the limitations of our tool, we manually inspected a sample of the tokens missed in the validation. This analysis reveals two main causes.
First, Huynh et al. derive their labels through manual inspection of contract source code. Consequently, some contracts are classified as honeypots based on the presence of restrictive or privileged functions (\eg functions that enable or disable trading). However, these functions are not necessarily invoked on-chain. Since \tool{} detects honeypots by observing their runtime behavior rather than their source code, it cannot identify contracts whose restrictive functionality is present but never activated. We note that this discrepancy does not necessarily represent a limitation of our approach, as such functions may also implement legitimate mechanisms. For example, temporarily disabling trading until the completion of a presale is a common and benign practice.
Second, \tool{} cannot detect blocklist-based honeypots. In these schemes, users are initially allowed to purchase the token but are later prevented from selling after the contract owner issues a separate transaction that blacklists the buyer. Because our evaluation simulates a single buy-then-sell sequence at a fixed block, it cannot capture restrictions that are introduced only after the purchase has taken place. 
However, such mechanisms are more complex, as they require dynamically adding addresses to a blocklist after token purchases, making them more costly and harder to deploy at scale.

Finally, the identification of 4,039 colluding addresses may be incomplete. Some addresses added to allowlists through unrecognized functions may have been missed, leading to an underestimation. However, given that these identified addresses account for 79.1\% of the total trading volume in \bt{} operations, we believe that our analysis captures a substantial fraction of the colluding activity.

\section{Related Work}

\mypara{\pd{}.}  
Prior work on \pd{} schemes has largely focused on centralized cryptocurrency markets, characterizing these events through the lens of price distortions~\cite{dhawan2023new}, long-term negative effects on targeted coins~\cite{clough2023pump}, and the particular vulnerability of low-capitalization assets~\cite{hamrick2019economics}. Several works propose automated detection models leveraging market data~\cite{kamps2018moon,xu2019anatomy,victor2019cryptocurrency,la2023doge} or social media signals~\cite{nghiem2021detecting}. Collectively, this body of work establishes that CEX-based \pd{} schemes follow predictable patterns of price manipulation at the expense of uninformed external investors.
Only recently have \pd{} schemes on DEXs received dedicated attention. Li et al.~\cite{li2021cryptocurrency} provide the first comparative analysis of \pd{} operations across both CEXs and DEXs, finding that early buyers on DEXs achieve rapid gains. Rath et al.~\cite{rath2024profiling} study 314 operations on PancakeSwap, categorizing participating addresses into bots, victims, and high-profit actors, providing a fine-grained view of participant roles within individual operations. Wu et al.~\cite{wu2025profit} propose PUMPWATCHER, a graph neural network framework that detects anomalous token transfer patterns associated with \pd{} activity, extending the detection from market signals to on-chain behavioral signatures.
Compared to prior work, our study provides a comprehensive characterization of \pd{} activity on DEXs. To the best of our knowledge, we are the first to document \bt{} schemes, analyze the tokens they target, and accurately quantify the profits generated while accounting for colluding addresses. This is particularly relevant because it shows that assumptions made in studies of \pd{} schemes on CEXs do not necessarily hold in the DEX setting, highlighting the need for dedicated analysis methods.

\mypara{Honeypot tokens.}
Torres et al.~\cite{torres2019art} introduced the concept of honeypot smart contracts, defined as contracts that appear exploitable but are deliberately designed to lock the funds of anyone attempting to exploit them. The honeypot tokens we study differ in their implementation and target: they modify standard ERC-20 functionality through custom functions that alter trading behavior and trap investor funds.
Chen et al.~\cite{chen2019tokenscope} propose TokenScope, which detects inconsistent ERC-20 token behavior by analyzing past transaction traces. Zhou et al.~\cite{zhou2025no} adopt a similar static analysis approach. Both methods are complementary to ours: while they identify whether a token exhibits suspicious behavior, they cannot determine whether honeypot restrictions were active at any specific point in time, a distinction our dynamic simulation is designed to capture.
Gan et al.~\cite{gan2023trick} and Huynh et al.~\cite{huynh2025serial} are the most closely related works to ours, as they also rely on transaction simulation to detect malicious tokens. Gan et al. analyze honeypots on DEXs, propose a taxonomy of mechanisms, and evaluate their simulation-based detection method on 10,000 Uniswap pools, identifying 8,443 abnormal pools. Huynh et al. similarly propose a classification of implementation techniques and a simulation-based detection tool. While these works focus exclusively on detecting honeypot tokens, our work addresses a different question: not only whether a token is a honeypot, but how these schemes are orchestrated in practice and how victims are systematically lured into them, a dimension neglected by prior work. Moreover, we release our full implementation as open source, enabling reproducibility and future extensions.


\section{Conclusion and Future Work}
In this paper, we present a large-scale study of 3,677 \pd{} operations conducted on DEXs across Ethereum and BSC, discovering that these schemes are fundamentally more predatory than their centralized exchange counterparts. Rather than relying solely on artificial price inflation targeting external investors, organizers systematically deploy honeypot tokens that permanently trap funds from their own channel subscribers, exploiting both the trust of their followers and the lack of awareness of external participants.

Using a detection tool that reconstructs the exact blockchain state at the time of each operation, allowing us to replicate trades carried out by the investors, we find that 71.5\% of observed schemes involve honeypot tokens. We further uncover the systematic use of colluding addresses to emulate market activity and create the impression of consistently successful operations, with these addresses accounting for 79.1\% of total trading volume in \bt{} operations.
Our profit analysis reveals that these operations are highly profitable for organizers, who gain in 99.3\% of cases, while participants collectively incur losses exceeding \$7.2 million. Even \pd{} operations also impose significant losses on late investors, totaling over \$22 million.

Several directions remain open for further investigation, including extending the study to additional blockchains and emerging DEX platforms. Future work should also enhance the tool to detect more sophisticated anti-simulation mechanisms and apply it to honeypot tokens in a broader context.

\cleardoublepage

\bibliographystyle{plain}
\bibliography{biblio}




\cleardoublepage
\appendix
\section{Open Science} 


In accordance with open science principles, and to support transparency and reproducibility, 
we publicly release the data and code produced in this study at the following link:
\url{https://gitfront.io/r/anonymized/uLnW981AuB8J/pump-and-dump-dex/}.

The repository includes the dataset extracted from Telegram (Sec.~\ref{sec:dataset}). 
The dataset contains one entry for each detected event and reports information about the 
Telegram channel coordinating the operation, the address, name, and symbol of the targeted 
token, the timestamp of the pump announcement, and the associated liquidity pool, decentralized 
exchange, and blockchain. This information can be paired with publicly accessible on-chain 
data, accessible via standard RPC endpoints or blockchain explorers such as Etherscan,
to fully reproduce our analysis. Beyond replication, the dataset can also support other 
research tasks, such as developing tools for prediction, real-time detection, or exploratory 
analysis of \pd{} activity on DEXs.

The repository also includes the source code of the tool used to perform transaction simulation  (Sec.~\ref{sec:detection_honeypot_tokens}). 
We release it as a standalone, reusable component that can be applied to arbitrary token addresses, independently 
of our dataset. This allows other researchers and practitioners to detect honeypot tokens on any DEX, enabling applications such as real-time scam detection, wallet 
protection tools, and large-scale ecosystem auditing.

Moreover, we release the source code of representative real-world smart contracts that 
implement honeypot mechanisms (Sec.~\ref{subsec:tool_results}). These contracts are 
manually selected examples drawn from our dataset, chosen to illustrate the concrete 
implementation techniques employed by malicious token deployers across the different honeypot 
categories we identify. They provide a ready-made reference for studying these patterns and 
can serve as a foundation for developing and benchmarking future detection approaches.

Finally, we release the function signatures used to identify colluding addresses 
(Sec.~\ref{subsec:detection_colluding_addresses}). Beyond reproducing our analysis, this 
dataset provides a reusable resource for identifying allowlist-based transfer restrictions 
across any token contract, lowering the barrier for future studies of coordinated trading 
behavior on DEXs.

\section{Ethical Considerations} 

Our analysis involves data collection from Telegram channels used to coordinate \pd{} operations. This process was conducted in strict accordance with Telegram’s Terms of Service~\cite{telegram_tos}, which permit the collection of publicly available chat data. Adhering to the GDPR principle of data minimization~\cite{gdpr_minimization}, we restricted our collection to messages specifically related to \pd{} announcements and relevant channel metadata. Furthermore, we intentionally avoided collecting personal identifiers, such as usernames and phone numbers, to ensure that no unnecessary or sensitive personal data was captured.

The remainder of the data used in this study was sourced from open and accessible public ledgers, namely the Ethereum and BNB Smart Chain (BSC) blockchains. All data retrieved from these sources is inherently public and pseudonymous; at no point did we attempt to deanonymize any user or entity.

To supplement this data, we occasionally utilized blockchain explorers (\ie Etherscan, BscScan) and Honeypot.is and GoPlusSecurity tools. Data collection from these sources was conducted responsibly: we avoided scraping and instead utilized official APIs, strictly adhering to the rate limits and usage policies provided by these services. 

All collected data pertains exclusively to smart contracts, token transactions, and market dynamics. Consequently, this study did not involve interaction with human subjects or the collection of Personally Identifiable Information (PII).

\subsection*{Positive impact on stakeholders}
\textbf{AMM investors.} For AMM investors, this research provides timely and practical insights into ongoing risks in decentralized markets. First, it offers an informative contribution by documenting that \pd{} schemes remain active and continue to target AMM-based markets. Second, our analysis highlights \bt{} operations that are even more dangerous, as individuals who are caught into these schemes face a high risk of being unable to sell the tokens they acquire. Disseminating this knowledge represents an important first step toward educating investors about the specific risks associated with trading activity in decentralized environments and reinforcing the importance of conducting due diligence before investing.

\textbf{AMMs and AMM-related platforms.} This work is also relevant to AMM platforms, token information aggregators, and trading interface providers. Our detection tool offers a practical and deployable mechanism for identifying honeypot tokens before investors commit funds. Crucially, because the tool is open-source and requires no proprietary infrastructure beyond access to an archive node, it can be integrated directly into DEX interfaces, blockchain explorers, and token information platforms. Platforms and interfaces that mediate user interaction with DEXs could leverage the tool to surface risk warnings at the moment of token discovery or apply additional scrutiny to newly listed assets. Such measures would help reduce user losses, preserve trust in decentralized trading infrastructure, and demonstrate proactive efforts to improve market integrity without compromising permissionless access.

\textbf{Regulators.} For regulators, this research provides a technical view of how \pd{} schemes operate on DEXs. As regulatory frameworks for cryptocurrencies become increasingly important, understanding these manipulations is essential to ensure that regulatory measures effectively address real-world abuses on decentralized markets abuses rather than relying on assumptions derived from centralized markets.

\subsection*{Negative impact on stakeholders}
\textbf{Malicious actors.}
Although this research aims to improve the mitigation of the phenomenon, it may also have unintended negative effects. By clearly documenting how \pd{} and \bt{} schemes operate on DEXs, the work could lower the barrier for malicious actors to replicate or refine these techniques. Less experienced attackers may learn which mechanisms are effective, while more sophisticated ones may adapt their strategies to evade detection methods. In addition, increased awareness of profitable manipulation patterns could temporarily incentivize copycat behavior before mitigation techniques are widely deployed.

\subsection*{Mitigation and Decision to Publish}
Although revealing the mechanics of \bt{} schemes could allow some adaptation by malicious actors, the benefits of this research far outweigh the risks. These techniques exploit well-known smart contract features already in use, and hiding them would mainly harm users, platforms, and regulators. In contrast, systematic documentation enables earlier detection, informed warnings, and practical mitigation strategies. Moreover, increasing transparency reduces information asymmetry which is the key factor that allows these schemes to succeed, thereby reducing the effectiveness of the scam. As defensive tooling and platform-level protections mature, the window of opportunity for misuse narrows, making the long-term impact of disclosure overwhelmingly positive.

\clearpage

\section{Appendix}

\begin{figure} 
    \centering
    \includegraphics[scale=0.28]{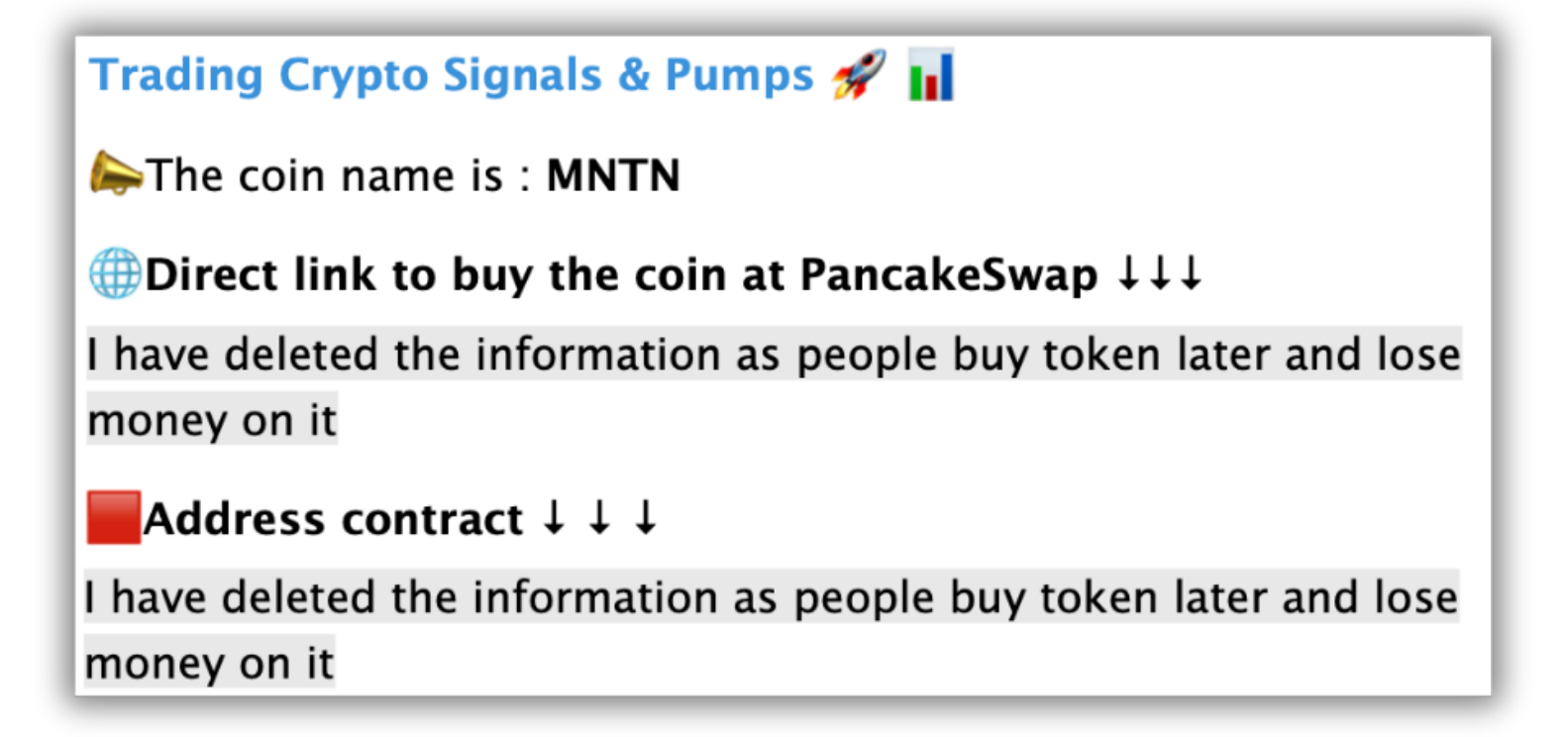}
    \caption{Telegram message with removed information.}
    \label{fig:telegram_message_obfuscated}
\end{figure}

\begin{table}
   \caption{List of words used to perform the search.}
    \label{tab:pump_and_dump_keywords}
    \centering{
        \begin{tabular}{l}
          \toprule
          \textbf{Keywords} \\
          \midrule
          Pump, Crypto Pump, Pump and Dump, Pump Signal, \\ 
          Pump Crypto Signal, Token Pump, Coin Pump, \\ 
          DEX Pump, Liquidity Pool Pump, DeFi Pump, P\&D \\
            \bottomrule
        \end{tabular}
        
        }
\end{table}

\subsection{Implementation analysis}
\label{appendix:implementation_anal}
To understand how the detected restrictions are implemented, we inspect representative tokens among those for which we recovered verified
source code during the dataset construction process described in
Section~\ref{sec:dataset}. This analysis provides concrete examples of
the code-level mechanisms underlying the runtime behaviors identified
by \tool{}.

For representative tokens exhibiting immediate sell reversion, we
observe checks within the token's \texttt{transferFrom} function that verify whether the seller is included in an \textit{allowlist}. If the seller is not present in the allowlist, a \texttt{require} statement causes the transaction to revert. This allows privileged addresses to sell the token while preventing ordinary buyers from doing so. We find the allowlist is often already populated with hard-coded addresses at deployment
time.

Timed sell reversion relies on a similar but more sophisticated
mechanism. In the inspected contracts, a Solidity \texttt{mapping}
stores the purchase timestamp associated with each buyer. When an
address attempts to sell, the contract computes the time elapsed since
its last purchase and reverts if it exceeds a predefined threshold.
This allows an immediate buy-and-sell simulation to succeed while
preventing sales after a delay that is more representative of human
trading.

Fee-based honeypots implement the fee within the token's transfer
logic. In the inspected contracts, buy-side fees are applied in the
\texttt{transfer} function, while analogous checks are applied to sell
transactions through the \texttt{transferFrom} function.

Finally, all tokens exhibiting approval reversion were promoted
exclusively by the \emph{BIG PUMP CRYPTO INVEST TRADING} Telegram
channel. This concentration suggests that the same implementation, or
closely related implementations, may have been reused across multiple
operations. However, none of the corresponding contracts has verified
source code, preventing us from determining the underlying mechanism.
The observed runtime behavior is consistent with a non-standard
\texttt{approve} function that reverts when the trader attempts to
authorize the DEX router, but we treat this explanation as a hypothesis.

To support reproducibility, we publicly release the complete source code of representative real-world tokens for each honeypot mechanism for which verified source code is available~\cite{additional_material}.

\begin{figure}
    \centering
    \includegraphics[scale=0.35]{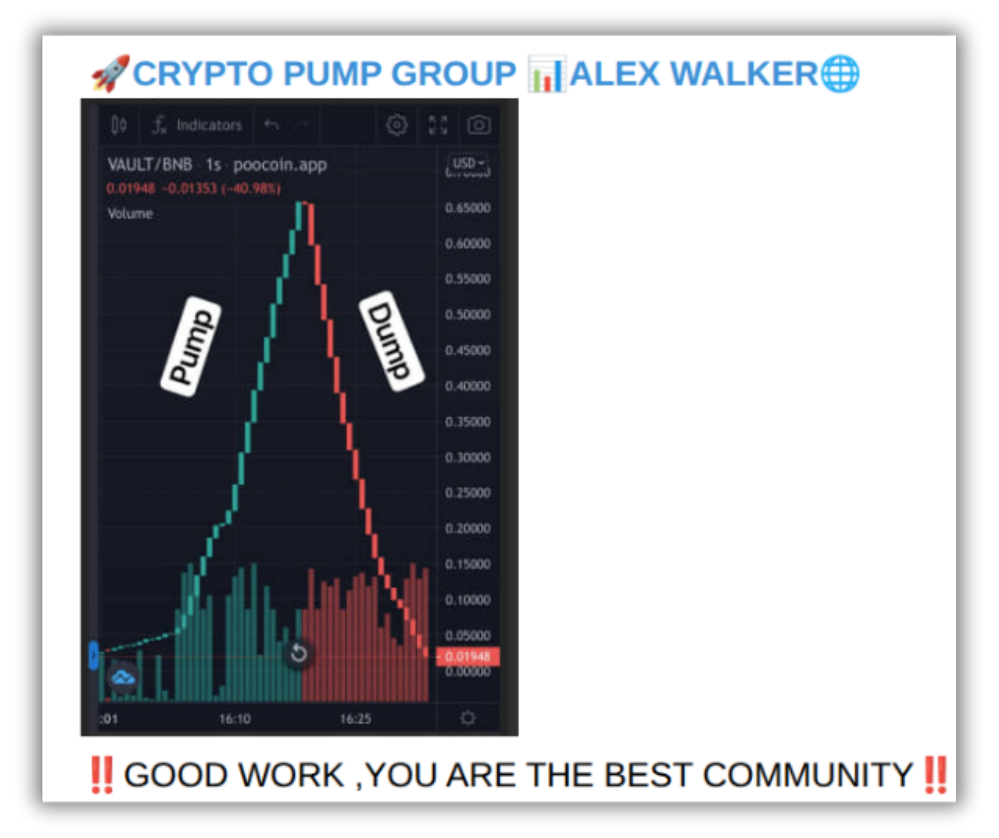}
    \caption{Telegram message claiming the operation as successful. The shared chart resembles the price pattern of a conventional \pd{} operation, but the observed price dynamics are almost entirely driven by addresses colluding with the token creator.}
    \label{fig:pump_successfull_message}
\end{figure}




%


\subsection{Known service dataset}
\label{subsec:known-service-dataset}
Flow fund tracking stops when funds reach addresses controlled by entities that aggregate assets from multiple users, such as CEX wallets (\eg Binance, Coinbase, Kraken), custodial services, or popular mixing services. These services act as aggregation and distribution points for transactions from many users, and any path that terminates within them is considered a taint boundary. 
To implement this feature, we build a \textit{Known Services Dataset}  that contains the addresses of services where the tainting procedure should stop. 
We build this dataset by aggregating information from multiple reliable sources, including labeled addresses from Etherscan and BSCScan~\cite{etherscan-labels}, tags from Dune Analytics~\cite{dune-labels}, and wallet holdings from DeFiLlama’s~\cite{defilama} transparency dashboard, which lists the 50 largest centralized exchanges. In total, this dataset contains 30,457 known service addresses on Ethereum and BSC.

\begin{figure}   
    \centering
    \includegraphics[scale=0.24]{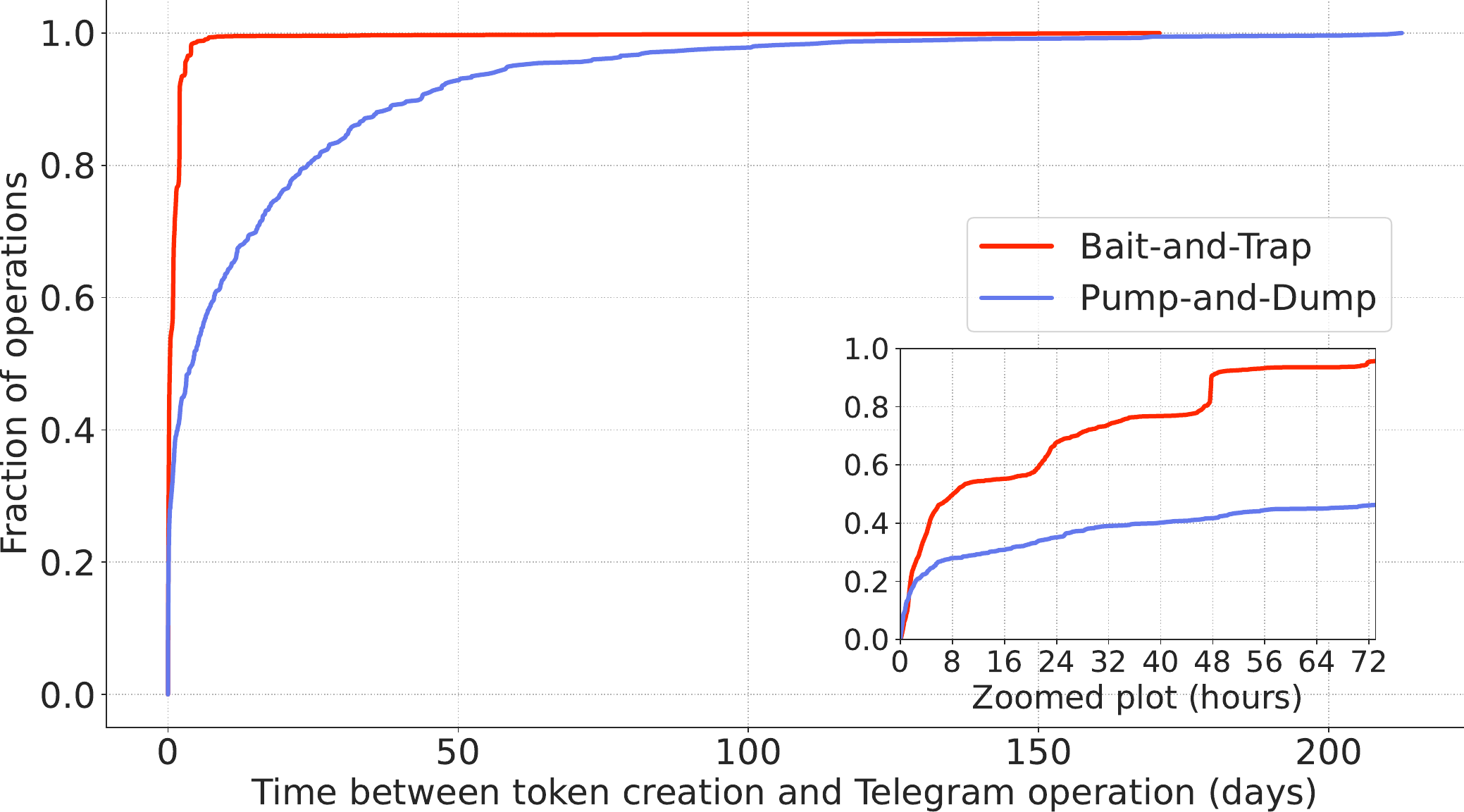}
    \caption{Time elapsed between the creation of the target token and the corresponding operation.}
    \label{fig:cdf_distance_token_creation_pump}
\end{figure}

\end{document}